\documentstyle[12pt,aaspp]{article}
\def\hi{\noindent \hangindent=2.5em}

\begin{document}

\title{The Properties of Poor Groups of Galaxies: II. X-ray and Optical Comparisons}
\author{John S. Mulchaey\altaffilmark{1} and Ann I. Zabludoff\altaffilmark{1,2}
\altaffiltext{1}{Observatories of the
Carnegie Institution of Washington, 813 Santa Barbara St., Pasadena,
CA 91101, E-mail: mulchaey@pegasus.ociw.edu}
\altaffiltext{2}{UCO/Lick Observatory and Board of Astronomy and
Astrophysics, University of California at Santa Cruz, Santa Cruz, CA,
95064, E-mail: aiz@ucolick.org}}

\begin{abstract}

We use ROSAT PSPC data to study the X-ray properties of a sample of twelve
poor groups that have extensive membership information (Zabludoff \& Mulchaey 1997; Paper I). 
Diffuse X-ray emission is detected in nine of these groups.
In all but one of the X-ray detected groups, the X-ray emission is centered on a
luminous elliptical galaxy. 
Fits to the surface 
brightness profiles of the X-ray emission suggest the presence of {\it two} 
X-ray components in these groups. 
The first component is centered on the 
central elliptical galaxy and is extended on scales of 20--40 h$_{\rm 100}$$^{-1}$ kpc.
The location and extent of this component, combined with its X-ray temperature
($\sim$ 0.7--0.9 keV)
and luminosity ($\sim$ 10$^{41-42}$ h$_{\rm 100}$$^{-2}$ erg s$^{-1}$), 
favor an origin in the interstellar medium
of the central galaxy. Alternatively, the central component may be the result of a 
large-scale cooling flow. 

The second X-ray component is
detected out to a radius of at least $\sim$ 100--300 h$_{\rm 100}$$^{-1}$ kpc. 
This component follows the same relationships found among the
X-ray temperature (T), X-ray luminosity (L$_{\rm X}$) and optical velocity dispersion($\sigma$$_{\rm r}$) 
of rich clusters. This result
 suggests that the X-ray
detected groups are low-mass versions of clusters and that the 
extended gas component can properly be called the intragroup medium, in analogy 
to the intracluster medium in clusters. 
The failure to detect an intragroup medium in the three groups with very low
velocity dispersions is consistent with their predicted X-ray luminosities and temperatures
based on the relationships derived for clusters and X-ray detected groups.
The best-fit value of $\beta$ derived from the
$\sigma$$_{\rm r}$-T relationship for groups and clusters is $\sim$ 0.99$\pm{0.08}$,
implying that the galaxies and hot gas trace the same potential with 
equal energy per unit mass and that the groups are dynamically relaxed. 

We also find a
trend for the position angle of the 
optical light in the central elliptical galaxy to align with the 
position angle of the large-scale X-ray emission.
This trend is consistent with that found for some rich clusters containing
cD galaxies (Rhee, van Haarlem \& Katgert 1992; Sarazin et al. 1995;
Allen et al. 1995).
The alignment of the central galaxy with the extended
X-ray emission suggests that the formation 
and/or evolution of the central galaxy is linked to the shape of the global group
potential. One possible scenario is that the central galaxy formed via galaxy-galaxy
mergers early in the lifetime of the group and has not been subject to significant 
dynamical evolution recently.

\end{abstract}


\section{Introduction}

Because most galaxies occur in small groups, understanding the physical
nature of these systems is critical for cosmology. An outstanding question
is whether poor groups are simply low-mass versions of richer clusters or physically
different systems. Establishing the nature of groups has proved difficult because these 
systems typically contain only three or four bright galaxies. Thus, it is not
even known if the majority of the cataloged groups are real, bound systems or
if they are simply chance superpositions (e.g., Ramella et al. 1989; 
Hernquist et al. 1995).

The presence of diffuse X-ray emission provides evidence for a 
common potential in some poor groups (e.g., Mulchaey et al. 1993, Ponman \& Bertram 
1993, David et al. 1994, Ebeling et al. 1995, Pildis et al. 1995, Henry et al. 1995,
Mulchaey et al. 1996a, Ponman et al. 1996, Burns et al. 1996).
Many ROSAT studies suggest  there is a correlation between the morphological
composition of the group and the existence of X-ray emission (e.g., Ebeling et al. 1995,
Pildis et al. 1995, Henry et al. 1995, Mulchaey et al. 1996a), providing further 
evidence for the reality of the X-ray detected groups. 
 However, it is 
still possible that the X-ray gas is merely a projection of unbound gas in filaments
along the line-of-sight (Hernquist et al. 1995).

Perhaps the best way to establish the physical reality of groups is to extend the kinematic
properties of these systems to include much fainter members. We have recently
completed a fiber spectroscopy study of a sample of twelve poor groups (Zabludoff \& 
Mulchaey 1997; Paper I) and have shown that the X-ray detected groups are indeed 
bound systems. Even if the reality of some groups has been demonstrated, there
is still controversy about the origin of the X-ray emission in these systems.
For example, both Dell'Antonio et al. (1994) and 
Mahdavi et al. (1997) find a much flatter relationship  between X-ray luminosity
and velocity dispersion ($\sigma$$_{\rm r}$) for groups than is found for rich clusters. 
These authors argue 
that the X-ray emission in groups
 is dominated by emission from individual galaxies and not from a global
group potential. In contrast, Ponman et al. (1996) remove the X-ray emission from individual
galaxies and find
reasonable agreement between the L$_{\rm X}$--$\sigma$$_{\rm r}$ relationship for
 Hickson Compact
Groups (HCGs) and rich clusters. 
Henry et al. (1995) and Burns et al. (1996) also find that the
X-ray luminosity function is a natural extension of the relationship for
clusters, implying a similar physical mechanism for the X-ray emission. However, both the
Henry et al. (1995) and Burns et al. (1996) studies
studies 
only include the most X-ray luminous groups and not the more typical, low luminosity
 systems where the
differences in the L$_{\rm X}$--$\sigma$$_{\rm r}$ relationship have been reported.

Some of the uncertainty in the L$_{\rm X}$--$\sigma$$_{\rm r}$ relationship and in other 
comparisons of optical and X-ray properties may result from poorly determined
optical properties. For most groups, velocity dispersions have
been estimated from as few as three or four galaxy velocity measurements. However,
in Paper I we showed
that calculating $\sigma$$_{\rm r}$ from
 only a few galaxies can result in large uncertainties in the
measurement, often underestimating $\sigma$$_{\rm r}$ by a factor of 1.5 or more. Thus, the 
values of $\sigma$$_{\rm r}$ used in many previous
 comparisons of the optical and X-ray properties of groups
may be flawed.

The X-ray observations of poor groups are also subject to large uncertainties that in general 
do not plague similar measurements in rich clusters. For example, 
diffuse emission from the global potential 
dominates the cluster X-ray emission. However, in poor groups, the contribution
of X-rays from individual galaxies may be substantial.
 The presence of luminous X-ray emission
in poor groups is found predominantly in systems with a central
elliptical galaxy (e.g., Mulchaey et al. 1996a). The expected X-ray emission from such
a galaxy can be comparable to the diffuse, extended emission observed in many groups.
Thus, it is not clear whether the observed emission is associated with an extended
halo of the central galaxy (e.g., Trinchieri et al. 1997) or represents hot gas in the
group's global potential. If the observed emission is dominated by a component from
the central galaxy's interstellar medium, comparisons with global group properties such as
velocity dispersion may not be particularly meaningful.

In Paper I, we showed that X-ray detected groups typically contain at least
$\sim$ 20-50 members to absolute magnitudes of
 M$_{\rm B}$ $\sim$ -14 to -16 + 5 log$_{\rm 10}$h$_{\rm 100}$.
With this many galaxy velocities available, we are able to calculate robust
velocity dispersions for poor groups. 
Here we use deep
ROSAT observations of the twelve groups studied in Paper I to examine the 
nature of the X-ray emission in these systems. In the next section, we 
describe the ROSAT observations and data reduction techniques. In \S3, we study the spatial
distribution of the X-ray emission and discuss the evidence for two X-ray
components in these groups. Spectral analysis of the ROSAT data, including temperature and
luminosity measurements, are presented in \S4. Comparisons of the derived optical 
and X-ray properties of groups are given in \S5, and our conclusions are summarized
in \S6.
All distance-dependent quantities in this paper 
are calculated assuming H$_{\rm o}$ = 100 km s$^{-1}$ Mpc$^{-1}$ (h=H$_{\rm o}$/100).

\section{ROSAT Observations and Data Reduction}

ROSAT PSPC observations were extracted from the archive for the group
sample given in Table 1. In all cases, the PSPC observation is centered on
the group. 
To search for diffuse gas in these systems,
 we followed the reduction method outlined in 
Mulchaey et al. (1996a). Times of high background are
identified and excluded by discarding any data taken when the Master Veto rate
is greater than 170 counts s$^{-1}$. The final exposure time used in 
each case is given in Table 1. 
For each group four images are created
corresponding to the energy bands above 0.5 keV described in Snowden et al. (1994).
These images are corrected for vignetting using energy-dependent exposure 
maps.  The four images are then co-added to produce a 0.5--2 keV image. 

Before the X-ray emission from the group can be studied, emission
from {\it unrelated} sources must be removed. 
We find these sources using the task \lq\lq DETECT\rq\rq \
in the Extended Object and X-ray Background Analysis software (Snowden 1994),
which identifies `point sources' in the field. In the 
context of this discussion, `point source' refers to a source that appears point-like
at the resolution of the ROSAT PSPC (FWHM $\sim$ 30$''$ corresponding to
$\sim$ 10 h$^{-1}$ kpc for our most distant group; the resolution is considerably 
worse for sources far off-axis). 
Emission from point sources is then removed by excluding a circular
 region 
around each source with a radius 1.5 times the radius that encircles 
90\% of the source flux. This exclusion radius corresponds to 1.5$'$ for 
sources on-axis (Hassinger et al. 1992). To verify that the majority of these point sources
are truly unrelated to the group, we compare the positions of the identified 
point sources with the positions of the known group members identified in Paper I.
For most (9 of 12) of the groups, only one or two of the group members 
are X-ray sources. The group with the most associated sources
is HCG 90, where four of the galaxies are X-ray emitters.
However, two of these galaxies are Seyferts, so the X-ray emission is probably
not related to the gravitational potential 
of the group or the interstellar medium
 of the host galaxies. Even in this extreme case, more than 
90\% of the point sources identified appear to be unrelated to the group. Given the
high rate of `contaminating' sources in these fields,
we exclude all point sources found by \lq\lq DETECT\rq\rq \ (except the 
central galaxy emission; see below).

The resulting (unsmoothed) image is then examined for evidence of diffuse emission.
In 9 of the 12 fields, a diffuse component is clearly visible. To our
knowledge, the 
diffuse emission detection in the NGC 4325 group has not
previously been reported. All the other positive detections are
documented in the literature (NGC 533, NGC 5129 and NGC 5846: Mulchaey et al. 1996a,
NGC 741: Doe et al. 1995, NGC 2563: Trinchieri, Fabbiano \& Kim 1997; 
HCG 42 and HCG 90: Ponman et al. 1996, 
HCG 62: Ponman \& Bertram 1993, Pildis et al. 1995).
Diffuse emission is also found $\sim$ 14$'$ north of the NGC 7582 group 
center. However, this emission is coincident with the background cluster Abell S1111 and
thus almost certainly originates in the rich cluster, so we count this system among
our non-X-ray detected groups.
In the 
NGC 491 and NGC 664 fields, no diffuse emission is detected.

\section{Spatial Analysis of the Diffuse Gas}

In Figure 1,
we overlay the contours for the diffuse X-ray emission on the 
STScI Digitized Sky Survey images of the group fields.
With the exception of HCG 90,
 the X-ray emission is always peaked within 5--10 h$^{-1}$ kpc of 
a giant elliptical galaxy, which is the 
brightest group galaxy (BGG). 
In Paper I, we showed that the BGG
lies near the kinematic and projected spatial center
of the group, suggesting it is
coincident with the core of the group potential (see \S3.4 of Paper I). 
For this reason, it may be proper to include the X-ray emission 
at the location of the BGG as part of the intragroup
gas and we have included this emission in Figure 1 (except for HCG 90).
All other emission due to point sources in the field has been 
removed.

For the groups with a detected diffuse component, an azimuthally-averaged surface brightness profile 
is constructed. The inner bin in each profile corresponds to 
a physical radius of $\sim$ 3 h$^{-1}$ kpc. We model the surface brightness profiles
using a modified King function: 

\centerline{S(R)=S$_{\rm o}$ 
(1.0 + (R/R$_{\rm core}$)$^2$)$^{-3\beta + 0.5}$} 

\noindent{with a constant background included in the fit. The models are convolved
with the 1 keV PSPC point spread function, and then fit to the data with
S$_{\rm o}$, R$_{\rm core}$, $\beta$ and the constant background as free parameters.
 For all but one of the groups with a BGG, the fits with a single model are unacceptable
(i.e., reduced chi-squared values $>$ 2.0).
In general, there is a shoulder in the profile at $\sim$ 2-3$'$, which suggests the presence 
of two X-ray components.

To test for the presence of two components, we fit
the surface brightness profiles with two King models.
Our fitting method is as follows. First, we fit the surface brightness profile 
beyond 5$'$ with a single King model and a constant for the background.
A single
King model provides a good fit to the extended component (see Table 2). 
Next, we fit the entire profile with the two component model, 
fixing the shape (i.e., $\beta$ and R$_{\rm core}$) of the extended component 
to the values found in the fit beyond 5$'$. To verify that this fitting method produces
robust results, we varied the inner radius over which the extended component was originally
fit from 3$'$ up to 10$'$. While the 
best fit parameters for the extended component were found to vary somewhat with the 
choice of the inner radius, they are consistent within the errors with those
determined in the initial fits (i.e., using an inner radius of 5$'$). Thus, although
an inner radius of 5$'$ corresponds to a different physical scale for each group, 
we do not believe this choice strongly effects our final fit results.
The parameters of the 
core component in the two King model fits are relatively insensitive to the 
adopted model for the extended gas.}

In general, the two component models provide a very good fit to the data and a
significant improvement over the single King model
(Figure 2 and Table 2). 
The success of this model argues for the presence of
two X-ray components in each of these groups. One possibility is that 
the X-ray emission from the core component originates in the 
interstellar medium of the central galaxy (the BGG), while the extended
component corresponds to gas in the global group potential (i.e., an 
intragroup medium). Ibeke et al. (1996) have suggested a similar model for
the poor cluster Fornax, attributing the sharp upturn in the surface brightness profile
at small radii to emission from the central galaxy and the extended component
to emission from the intracluster medium.
The X-ray properties of the two components in poor groups 
appear to be consistent with a similar interpretation.
We examine this idea further in \S5.

An interesting consequence of the two component models is that the derived slopes 
of the King profiles (i.e., $\beta$) for both 
components tend to be much steeper than is implied from a single
King model fit to the observed emission. Mathematically this behavior
is expected when a narrow/brighter beta model and a wider/fainter
beta model are added together (Makishima 1995).
 The larger values of $\beta$ implied from the two component
fits has several important implications for groups. In a 
model where both the gas and galaxies are isothermal and in equilibrium, $\beta$ is 
simply the square of the ratio of the galaxy to gas velocity dispersion:

\centerline{$\beta$ = $\sigma$$_{\rm r}$$^2$$\mu$m$_{\rm p}$/ kT}
 
\noindent{where $\mu$ is the mean molecular weight in amu, m$_{\rm p}$ is
the mass of the proton, $\sigma$$_{\rm r}$ is the one-dimensional velocity dispersion and 
T is the temperature of the gas. 
Previous X-ray studies
of groups suggested low values of $\beta$ for the extended gas, 
implying that the energy per unit
mass is higher in the gas than in the galaxies. However, our two component fits 
are consistent within the errors with $\beta$ $\sim$ 1 for 
the extended component in many cases. Thus, significant heating of
the gas by non-gravitational processes may not be required in these groups.
Furthermore, many of the derived physical parameters for groups such as the
total group mass and the fraction of mass in hot gas depend on the
value of $\beta$ assumed. For example,
for a fixed X-ray luminosity, 
higher values of $\beta$ lead to significantly lower gas masses. Thus, if our
$\beta$ values are correct, the baryonic mass in hot gas may be less
than what has been inferred in some previous studies. The $\beta$ parameter
can also be derived independently from direct measurements of $\sigma$$_{\rm r}$ and 
of temperature 
as we discuss in \S5. 

While the evidence for two X-ray components in these poor groups is compelling,
the parameters derived from the King model fits depend strongly on the 
assumptions made. In general, most of the derived X-ray properties of
groups are poorly determined compared with rich clusters, which have a higher surface 
brightness and a lower level of relative contamination from individual galaxies.
Still, many of the effects that we find may also be relevant to the X-ray properties
of richer systems. For example, in analogy to the 
X-ray detected groups, we might expect a second X-ray component to be associated with the
central galaxy in cD-dominated clusters. Although this component might be more difficult to 
discern against the more luminous intracluster medium, it would probably have a non-negligible
effect on the surface brightness profile and derived physical quantities like $\beta$.

\section{Spectral Analysis}

Given the results of the surface brightness profiles, we
have extracted three spectra
for most of the X-ray detected groups: 
i) one for the central component, ii) one for the
extended gas component,
and iii) one with both the extended and central components included.
Because it is not possible to spectroscopically isolate the central emission
from the extended emission with the ROSAT PSPC data, 
we assume that 
the flux in the inner 1.5$'$ is dominated by the central emission and that the 
rest of the
 flux is from the 
extended component. The only
exception is the closest of the X-ray detected groups NGC 5846, where 
we assume that all the flux within 3$'$ of the center is 
from the central component (see Figure 2).
For the NGC 5129 group there are insufficient counts in the central component to study
it spectroscopically, so we have extracted a spectrum for the extended gas only.
In the case of HCG 90, where the diffuse gas is not centered
on a particular galaxy, we also extract only a diffuse spectrum.

 The maximum extent of each spectral extraction is given in column 3 of Table 3. For five
of the nine 
groups with an extended
 component, the maximum extraction radius corresponds to 0.3 h$^{-1}$
 Mpc. 
This radius is chosen because it is the smallest radius within which 
there are enough galaxy velocity measurements so that we can obtain 
a reliable velocity dispersion for the groups (see Paper I).
However, for HCG 42, NGC 5846 and HCG 90, the diffuse X-ray emission 
is detected to only a fraction of 0.3 h$^{-1}$ Mpc. The spectra for these three groups were 
extracted 
at a radius where the surface brightness of the diffuse gas reaches 20\% of the determined
background level. 
Point sources
were excluded from each spectral extraction using the circular regions described in \S2. 
The background region in each case
was an annulus with an inner radius beyond the edge of the detectable emission.
Each spectrum was fit with two different 
spectral models. The first model was a Raymond-Smith plasma model with variable abundances
and fixed absorbing column. The results of these fits are given in columns 4--6 of 
Table 3 (all quoted errors are for the 90\% confidence level).
In general, good fits to the data are found  with this model (i.e., reduced chi-squared less than 1.5).
Good fits are also obtained with a MEKA plasma model (Mewe, Gronenschild \& van den Oord 1985)
and variable abundances (see columns 7--9 of Table 3). 
An examination of the temperatures and abundances derived from
the two spectral models
 reveals good agreement between the models for the systems studied here. 
While the metal abundances implied tend to be very low, we caution the reader that this quantity
may not be reliably determined with the ROSAT PSPC
(cf. Bauer \& Bregman 1996) and so strong statements about the gas enrichment 
cannot be made with the current dataset. To determine the effects of the metallicities on 
the derived gas temperatures, we also fit the diffuse spectrum of each group with a 
Raymond-Smith model with the metallicity fixed at half solar (Table 3). With this 
assumed metallicity, the derived gas temperatures were typically $\sim$ 10\% higher, consistent
within the errors to the best-fit value in the model where the abundance was allowed to
vary.

 For three of the groups, the diffuse gas
spectrum has enough counts that a rough temperature profile can be
determined. In these cases, we have extracted spectra in several annuli, which were chosen to
contain $\sim$ 1000 counts in each bin. The innermost point in each temperature
profile corresponds to the emission centered on the central elliptical galaxy.
The profiles are given in Figure 3.
Temperature profiles have previously been published for all three groups
(NGC533 and NGC 2563: Trinchieri et al. 1997, HCG 62: Ponman \& Bertram 1993,
Pildis et al. 1995) and our results are consistent with these earlier studies. 
 For the NGC 533 group, our three temperature measurements suggest an isothermal
distribution out to the radius of 300 h$^{-1}$ kpc. Trinchieri et al. (1997) bin the data into
smaller annuli and find some indication that the gas temperature may rise slightly beyond
the central point, although the uncertainties in their derived temperatures are large
given the smaller number of counts in each bin.
In the NGC 2563
group, the profile is consistent with an isothermal distribution beyond
about 2$'$ ($\sim$ 30 h$^{-1}$ kpc). However, the gas in the central region is cooler than the rest
of the group by nearly 20\%. A similar effect is seen in HCG 62, where 
the temperature of the gas rises from the center out to 
a radius of $\sim$ 7$'$ ($\sim$ 90 h$^{-1}$ kpc) and then drops at large 
radii. 
Although we do not have enough counts to produce temperature profiles 
for the other groups, in at least one other case (NGC 741), 
the central component is significantly cooler than the extended gas (see Table 3).
The lower temperature of the gas in the inner regions of these groups
further supports the existence of a distinct component 
possibly associated with the central galaxy.
The thermally cooler X-ray gas near the centers of these groups is consistent
with the central galaxy being dynamically cooler (i.e., the velocity dispersion 
is lowest near the center of the group; see \S3.3 of Paper I).

The luminosity of the X-ray emitting gas in each group
 has been estimated from the Raymond-Smith
spectral model with the best fit abundance and the absorbing column fixed at the  
Galactic value. The X-ray luminosities depend on several assumptions.
First, some authors have adopted a 
{\it mixed emission} scenario for groups (e.g., Mahdavi et al. 1997),
where it is assumed that most of the group gas is associated with the potentials of individual
galaxies. In these cases, the authors have included the contribution from the 
galaxies in the total X-ray emission. Other authors have excluded the emission from
individual galaxies, 
preferring instead to measure the luminosity of the extended,
 diffuse gas only (e.g., Ponman et al. 1996). 
For all of the X-ray detected groups in our sample except HCG 90,
 the diffuse gas plus the central galaxy 
emission dominates the observed flux. Typically, these components account for 60--80\% of
the  raw counts after background subtraction (the flux of the extended component is 
typically 2--10 times that of the central component). 
The remaining X-ray emission is associated with 
`point sources'. 
 As described in \S2, more than 90\% of these `point sources' appear to 
be unrelated to the group as a whole or even to the individual group galaxies. In fact, for three of the 
X-ray detected groups, the only X-ray emitting member is the BGG.
Given that the majority of the `point source' flux is completely unrelated to the 
global group potential or to the group galaxies, we exclude it from the calculated luminosities.
However, since a fraction of the diffuse emission 
is lost when point sources are masked out, we have corrected the 
extended component luminosities
 for this \lq\lq missing\rq\rq \  
flux by replacing the masked regions with
interpolated values. This correction typically raises the diffuse luminosities by
$\sim$ 20\%. 

We have also assumed the best-fit metallicities in calculating 
the luminosities. Unfortunately, the X-ray luminosity is a strong function of 
the assumed metallicity and if our metallicities are lower by a factor of a few, the 
luminosities may be significantly overestimated (compare best-fit to 0.5 solar models 
in Table 3). 
The luminosities for the extended component, the central component and the 
combination of the extended + central components are given in Table 3. We
have estimated errors in the luminosity by varying the temperature and 
metallicity parameters over their 90\% confidence range (see Table 3).
For the groups where X-ray luminosities have been published in the literature
(e.g., Ponman et al. 1996; Trinchieri et al. 1997), there is good agreement between
our derived luminosities and the values found by previous authors. 

For the three groups where no diffuse emission is seen, we derive upper limits
on the X-ray luminosities using a method 
similar to that described in Mulchaey et al. (1996a). 
We first measure the net counts in a region 300 h$^{-1}$ kpc
in radius centered on each group.
The counts are then converted to flux assuming
a Raymond-Smith plasma model with temperature 1 keV and abundance 0.3 solar.
All point sources in the field have been masked out as described above.
Thus, the quoted luminosities correspond to upper limits for the {\it diffuse}
X-ray emission in these groups. The diffuse emission $\sim$ 14$'$ north of the 
NGC 7582 group has been included in the calculation of the luminosity for this system.
However, because this emission is almost certainly associated with the background cluster
Abell S1111, the true group luminosity is probably much lower.

\section{Discussion}

The X-ray analysis presented in the last two sections
argues strongly for the presence of two X-ray components in the X-ray detected groups:
one restricted to the inner few tens of kiloparsecs of the group and one on the 
hundreds of kiloparsec scale. The existence of these two components is 
supported by the surface brightness profiles of the X-ray emission and by differences
in the temperature of the gas between the inner and outer regions of the groups.
In this section, we consider the nature of these components further.

As suggested earlier, the central component may originate in the 
interstellar medium of the BGG.
The idea is supported by 
several observations. First, the location of the central component is always
consistent with the optical position of the BGG (within $\sim$ 5--10 h$^{-1}$ kpc,
 one PSPC resolution element at the distances of these groups). 
The X-ray properties of the central component are also consistent with those expected
for a luminous elliptical galaxy. For example, the core
radius of the central component
is in the range $\sim$ 4 to 20 h$^{-1}$ kpc, comparable to
what has been found in previous studies of
ellipticals (i.e., Forman, Jones \& Tucker 1985;
Thomas et al. 1986; Trinchieri, Fabbiano \& Canizares 1986;
Canizares, Fabbiano \& Trinchieri 1987; Buote \& Canizares 1994). 
The X-ray luminosity of the central component ($\sim$ 10$^{41-42}$ h$^{-2}$ erg s$^{-1}$) 
 is consistent with the X-ray luminosity of other
ellipticals of similar blue optical luminosity
(e.g., Fabbiano, Kim \& Trinchieri 1992; Eskridge et al. 1995). 
Finally, the temperature of the central component ($\sim$ 0.7--0.9 keV)
is in the range of temperatures found for elliptical galaxies with 
stellar velocity dispersions ($\sim$ 200--300 km s$^{-1}$) similar to the BGG 
(e.g., Matsumoto et al. 1997).
In summary, the location, extent, luminosity and temperature of the central 
component suggest that it is most likely associated with the BGG.

An alternative explanation for the central component is that it is the result of
a large-scale cooling flow. Some of the X-ray signatures of cooling flows in clusters
include sharply peaked surface brightness profiles, drops in the gas temperature near
the center of the cluster and excess X-ray absorption (cf. Fabian 1994). In addition,
many of the central galaxies in cooling flow clusters have unusual optical properties,
such as strong, low-ionization emission-lines and diffuse blue light (e.g., 
Heckman et al. 1989; Crawford \& Fabian 1992). The X-ray
detected groups also display some of these signatures. In particular, the sharp
upturn in the surface brightness profiles indicates that the gas
density
is rising steeply towards the center of groups, as expected 
in a cooling flow model. In contrast to expectations, however,
the gas temperature near the center of some groups
is comparable to the outer gas temperature (e.g., NGC 533).
We have also searched for optical emission-lines in the BGGs using our
fiber spectroscopy data, but no line emission is found.

 While these observations
do not necessarily rule out the cooling flow model,
we believe the X-ray and optical data may be more naturally explained if the 
central component originates internally in the interstellar medium of the BGG.
As described above, the X-ray properties of the central component are all consistent
with those expected for elliptical galaxies. Furthermore, the cooler temperature of
the central component in some groups is expected because the BGG tends to be dynamically
cooler than the group as a whole (i.e., the velocity dispersion profile of the group
also drops at the center; see \S3.3 of Paper I). Future X-ray observations 
should help distinguish between an external (i.e., cooling flow) and internal (i.e., BGG
interstellar medium) origin for the central component. Gas metallicity measurements,
in particular, should provide insight into this problem.

The extended X-ray component can be traced to at least a radius of 100--300 h$^{-1}$ kpc
in the X-ray detected groups. The large extent of this component suggests that it 
may be associated with the global group potential, much as
the hot, diffuse gas in rich clusters is associated with the deep
cluster potential well. 
We address this issue further by comparing our group sample
with the large cluster sample in 
Mushotzky \& Scharf (1997), which includes clusters at low and intermediate 
redshift with well-determined X-ray temperatures and luminosities.

Because the collision time-scale for gas in groups and clusters
is short compared to
the gas cooling time,
the temperature of the extended hot gas provides a 
tracer of the global potential.
The velocity dispersion of the system also provides a measure of
the potential. 
Our calculated velocity dispersions do not suffer from the
statistical uncertainties that have previously made this quantity unreliable for groups,
because we have measured velocities for at least 20--50 members in the
X-ray detected groups.
We plot the X-ray temperature versus the optical velocity dispersion for the group
and cluster samples in Figure 4. 
The best-fit parameters and
errors have been determined 
using the parametric bootstrap technique described in Lubin \& Bahcall (1993).
For the group sample, the best fit is:

\centerline{Log $\sigma$$_{\rm r}$ = (2.56$\pm{0.06}$) + (0.45$\pm{0.19}$) log T.}

\noindent{The fit to the entire sample of groups and clusters gives:}

\centerline{Log $\sigma$$_{\rm r}$ = (2.59$\pm{0.04}$) + (0.51$\pm{0.05}$) log T}

\noindent{which is in good agreement with both the relationship we find for groups and the 
relationship found by others for rich clusters
(e.g., Edge \& Stewart 1991; Lubin \& Bahcall 1993; Mushotzky \& Scharf 1997).}

As noted earlier, 
it is traditional to use the $\beta$ parameter in clusters to infer the ratio of the average 
energy per unit mass in the galaxies to the average energy per unit mass in the gas. 
For a simple
model where both the galaxies and the gas are 
isothermal and trace the same potential, $\beta$ = 1. The best-fit value of
$\beta$ derived from our fit to the $\sigma$-T relationship for the group and cluster sample is
$\beta$ = 0.99$\pm{0.08}$,
consistent with the isothermal model. 
The $\beta$ determined for the group
sample alone is also consistent with this value ($\beta$ = 1.02 $\pm{0.16}$) and 
with the $\beta$ value found from most of the surface brightness profiles (\S3).
This result conflicts with some earlier claims that most groups 
have $\beta$ $<$ 1 (e.g., Ponman et al. 1996). We suspect these differences result
from the ill-defined velocity dispersions used in previous studies. While a
velocity dispersion determined from only three or four velocity measurements
can either underestimate or overestimate the true group dispersion, a 
dispersion grossly in error is more likely to be underestimated (see Paper I).
This bias could explain the trend Ponman et al. (1996) find for low values of $\beta$.
The observation that the typical $\beta$ value for groups is $\sim$ 1 implies that
the cores of the X-ray detected groups are (on average) virialized. This conclusion is also supported
by the optical kinematics of these groups (see Paper I).

The X-ray luminosity provides
a measure of the total mass in hot gas. 
For rich clusters, there is a correlation 
between this quantity and the total mass of the system
(as measured by T or $\sigma$$_{\rm r}$: Edge \& Stewart 1991, David et al. 1993,
Mushotzky \& Scharf 1997).
 In Figure 5, we plot the X-ray luminosities of the hot gas
versus the gas temperature. Although there is considerable scatter in the group
sample alone, a similar scatter is found for the clusters. When the groups and 
clusters are considered together, a correlation is found.
The fit to the entire sample gives:

\centerline{Log L$_{\rm X}$ = (42.44$\pm{0.11}$)+log h$^{-2}$ + (2.79$\pm{0.14}$) log T.}

\noindent{Ponman et al. (1996) find a significant steepening in the L$_{\rm X}$--T 
relationship for groups and suggest this may be evidence for the importance of galactic
winds in low-mass systems. While we have too few points to determine a relationship for
the group sample alone, we find no evidence for a steeper relationship at low ($<$ 1 keV)
 temperatures. Our groups are consistent with the observed cluster relation.

There is also a trend between X-ray luminosity and velocity dispersion for
the group + cluster sample
(Figure 6 and Table 4), although once again, the scatter is considerable:

\centerline{Log L$_{\rm X}$ = (31.61$\pm{1.09}$) + log h$^{-2}$ + (4.29$\pm{0.37}$)log $\sigma$$_{\rm r}$.}

\noindent{We have not included the three non-X-ray detected groups
in our derivation of the L$_{\rm X}$-$\sigma$$_{\rm r}$ relationship. 
We note, however, that the upper limits on the X-ray luminosities of the 
three non-detected groups are consistent with the luminosities expected given their
low velocity dispersions (e.g., Mulchaey et al. 1996b; Ponman et al. 1996).}

Two recent studies have suggested significant flattening
of the L$_{\rm X}$-$\sigma$$_{\rm r}$ relationship for groups (Dell'Antonio et al. 1994;
Mahdavi et al. 1997). The dashed line in Figure 6 gives the best-fit relationship
Mahdavi et al. (1997) find for their sample of nine groups observed in the 
ROSAT all-sky survey. While most of our groups scatter around the 
midpoint of the Mahdavi et al. (1997)
relationship, the much steeper extrapolation of 
the cluster relationship also provides an adequate 
description of the groups in our sample (Figure 6). Ponman et al. (1996) also find that 
a sample of Hickson Compact Groups are consistent with the cluster extrapolation. 
Mahdavi et al. (1997) 
suggest that the differences between their derived slope and that of Ponman et al.
(1996) can be understood by two effects. First, the velocity dispersions used by
Ponman et al. (1996) were based on an average of four galaxies and thus might not 
provide a good estimate of the group potential.
Second, Ponman et al. (1996) have removed X-ray emission from individual
galaxies, which according to the Dell'Antonio et al. (1994) model may be the 
dominant contribution to the total group X-ray emission. 

Our sample should not suffer from the effects suggested by Mahdavi et al. (1997).
In fact, the quality of both our optical and X-ray data is much higher than that used in 
most previous studies. 
However, like
Ponman et al. (1996), we
derive a L$_{\rm X}$-$\sigma$$_{\rm r}$ relationship
for groups that is consistent with what has been found for clusters.
The difference between our result and that of Mahdavi et al. (1997) can probably
be explained by the different techniques used to measure the X-ray luminosity. 
By excluding emission from individual galaxies, we are studying the X-ray emission
associated with the global group potential. In contrast, Mahdavi et al. (1997)
are studying the combined emission from the individual galaxies and the intragroup
medium. Given the different quantities being compared, differences
in the derived L$_{\rm X}$-$\sigma$$_{\rm r}$ relation would be expected.
Furthermore, as groups tend not to span a large range in 
velocity dispersion or X-ray luminosity, it is
difficult to determine a reliable  L$_{\rm X}$-$\sigma$$_{\rm r}$ 
relationship for these systems alone from the limited number of groups with 
well-determined velocity dispersions and X-ray luminosities.
A firm determination of the L$_{\rm X}$-$\sigma$$_{\rm r}$ relationship for groups
will probably have to wait until many more groups have reliable X-ray and optical
measurements.
However, it does appear that when only the
intragroup medium component is considered in the measurement of 
X-ray luminosity (as in the present paper),
X-ray detected groups are consistent with the
extrapolation of the cluster L$_{\rm X}$-$\sigma$$_{\rm r}$ 
relationship.

The fact that the extended X-ray component in groups follows the same
L$_{\rm X}$-T-$\sigma$$_{\rm r}$ relationships
found for rich clusters of galaxies 
argues that these X-ray groups 
are low-mass extensions of the cluster phenomenon. 
Given this, the extended component can properly be thought of as the intragroup 
medium, in analogy to the intracluster medium in clusters. The nature of the 
non-X-ray detected groups is less clear. Based on the low velocity dispersions
of these groups, we would not expect to detect diffuse emission in the ROSAT
observations. Thus, the present X-ray observations do not allow us to determine
whether the non-X-ray-detected groups are bound systems or just chance 
superpositions of galaxies along the line-of-sight.

Further evidence of the physical similarities of the X-ray groups and rich clusters
may be found by considering the BGG-group interface. 
There is some evidence that the BGG \lq\lq knows about\rq\rq \
the extended group potential: there
is a clear tendency for the diffuse X-ray
emission to roughly align with the optical light of the galaxy in most cases. 
To quantify
this trend, we have fit the light distributions of the central galaxy and the intragroup
medium with ellipses (using the task `ellipse' in IRAF). 
The ability to determine a reliable position angle (PA) from an ellipse fit depends
on several factors. First, the PA is not a meaningful quantity when the ellipticity
of the X-ray emission is very small. This is often true in the inner few arcminutes of 
the group. Here we only consider a X-ray PA measurement reliable when the ellipticity 
is greater than 0.1. The ability to measure a PA is also a function of the signal-to-noise
ratio. For most of the groups, this restricts accurate X-ray PA measurements to radii less
than 10$'$ ($\sim$ 100-200 h$^{-1}$ kpc).
 Keeping these limits in mind, we plot the determined PA's of the 
X-ray emission as a function 
of radius in Figure 7 for the seven groups with sufficient data 
(meaningful X-ray PA's cannot be measured for the NGC 5846 group and 
HCG 90 because of low ellipticity and low signal-to-noise ratio, respectively).
Figure 7 reveals 
that the radial variations in the X-ray PA's of most of the groups are small (i.e.,
less than $\sim$ 20 degrees).
The X-ray isophotes are more complex in NGC 2563 and NGC 4325, possibly
indicating the presence of isophotal twisting.

We have also determined the optical
PA's of the central galaxies from CCD images in the B-band (see Table 5 and Figure
7). These measurements are typically made at much smaller radii than the X-ray
measurements.
As can be seen from Figure 7, 
the optical PA of the central galaxy and the PA of the diffuse X-ray emission often align to
within the errors of the measurements.
In five of the seven groups, the optical and X-ray PA's agree to better than 20 degrees.
The probability of this occuring randomly if the parent sample is uniform is less than
7 $\times$ 10$^{-3}$.
We conclude that there is a strong trend for the optical light of the
BGG and the global X-ray emission to align. 
A similar 
phenomenon has been seen in rich clusters containing cD galaxies
 (e.g., Rhee, van Haarlem \& Katgert 1992;
Sarazin et al. 1995; Allen et al. 1995), implying that the physical mechanisms responsible
for the formation of the BGG and cD galaxies may be similar.
The alignment of the central galaxy isophotes with the global group potential
(as probed by the X-ray emission) is consistent with the BGG forming via
galaxy mergers early in the 
lifetime of the group. 
The low velocity dispersions of
groups make them a likely site for such mergers. Mergers are
most probable during the initial formation of the group because at this point the
galaxies still retain large halos (Merritt 1985). The fact that the BGG is
at rest in the center of the group potential (Paper I) also suggests that the core of
the group has not experienced much dynamical evolution recently and that it is near
or at equilibrium.

\section{Conclusions}

We have presented ROSAT PSPC observations of twelve groups with detailed 
galaxy membership (Paper I).
Diffuse X-ray emission is found in nine groups.
The most luminous galaxy in each of the X-ray detected groups is an elliptical,
whose position is coincident with the peak of the X-ray emission in 
all but one case.
Surface brightness 
profiles of the X-ray emission strongly suggest the presence of two 
components in these groups: one on scales of $\sim$ 20--40 h$^{-1}$ kpc 
and one on much larger scales (of at least $\sim$ 100--300 h$^{-1}$ kpc).
The temperatures of the central and extended components are significantly
different in some groups, consistent with the interpretation that the components
are distinct.
The extent, temperature ($\sim$ 0.7--0.9 keV) and luminosity ($\sim$ 10$^{41-42}$ 
h$^{-2}$ erg s$^{-1}$) of the first component is consistent 
with that observed in elliptical galaxies in other environments,
suggesting that this component 
most likely originates in the interstellar medium of the central galaxy. 
Alternatively, the central component may be the result of a cooling flow.

The extended X-ray 
component follows the extrapolation of the relationships found 
among velocity dispersion, X-ray temperature and  X-ray
luminosity for rich clusters.
 This suggests X-ray detected poor groups can be thought of as scaled-down versions 
of clusters, with the extended X-ray component in groups representing the intragroup
medium, in analogy to the intracluster medium in clusters.
The best fit to the $\sigma$$_{\rm r}$-T relationship for X-ray groups and clusters
gives a mean $\beta$ value of 0.99$\pm{0.08}$, suggesting the galaxies and hot 
gas trace the same potential and that the energy per
unit mass in the gas and galaxies is equal. The values of $\beta$ derived independently
from fits to the surface brightness profiles are consistent with $\beta$ $\sim$ 1 
in many groups. The lower values of $\beta$ implied from other studies may be due to 
significant underestimates of the group velocity dispersion and to contamination
of the surface brightness profiles by the central galaxy X-ray emission. 

While the ROSAT data suggest the presence of two independent X-ray components
in these groups, it is clear that the formation and/or evolution of the central galaxy
is somehow linked to the extended group potential. In particular, there is a 
strong trend for the optical isophotes of the central galaxy to align with the 
X-ray emission isophotes on large scales. This result might be expected if the 
central galaxy formed via galaxy-galaxy mergers early in the lifetime of the 
group and has not been recently disturbed. The fact that the 
central galaxy is at rest in the center of the group's potential 
(Paper I) is consistent with this scenario. A similar phenomenon has been observed
in some clusters (e.g., Rhee, van Haarlem \& Katgert 1992; Sarazin et al. 1995;
Allen et al. 1995),
offering further evidence of the similarities between X-ray luminous,
poor
groups and rich clusters.

The results of this paper demonstrate the insight that can be gained by combining
detailed group membership information with quality X-ray observations. Still, there
are many outstanding questions that can be addressed with 
further work. Our sample contains
few non-X-ray detected or low temperature groups. We are in the process of extending 
our spectroscopy program to include more of these systems. Significant improvements 
in the X-ray observations can also be expected. Many groups have been observed with 
ASCA, which should help determine the metallicity of the intragroup medium, 
providing further constraints on the origin of this component. Higher
spatial resolution observations will also allow the central galaxy-group interface to
be studied in much more detail.

\acknowledgments

The authors would like to thank Lori Lubin for her assistance with the
parametric bootstrap technique, Richard Mushotzky and Caleb Scharf for
providing data from their cluster sample, 
Dennis Zaritsky for a careful reading of the manuscript 
 and the referee for useful comments that
improved this paper.
The authors also acknowledge valuable discussions with 
Mike Bolte, David Burstein, Julianne Dalcanton, David Davis, Megan Donahue, Lars Hernquist, Tod Lauer, 
Lori Lubin, Bill Mathews, Richard Mushotzky, Julio Navarro, Ian
Smail and Dennis Zaritsky. This research was made possible with the use of the
HEASARC and NED databases. JSM acknowledges partial support for this program 
from NASA grants NAG 5-2831 and NAG 5-3529 and
from a Carnegie postdoctoral fellowship. AIZ acknowledges support from the Carnegie
and Dudley Observatories, the AAS, NSF grant AST-95-29259, and NASA
grant HF-01087.01-96A.

\begin{figure}
\plotone{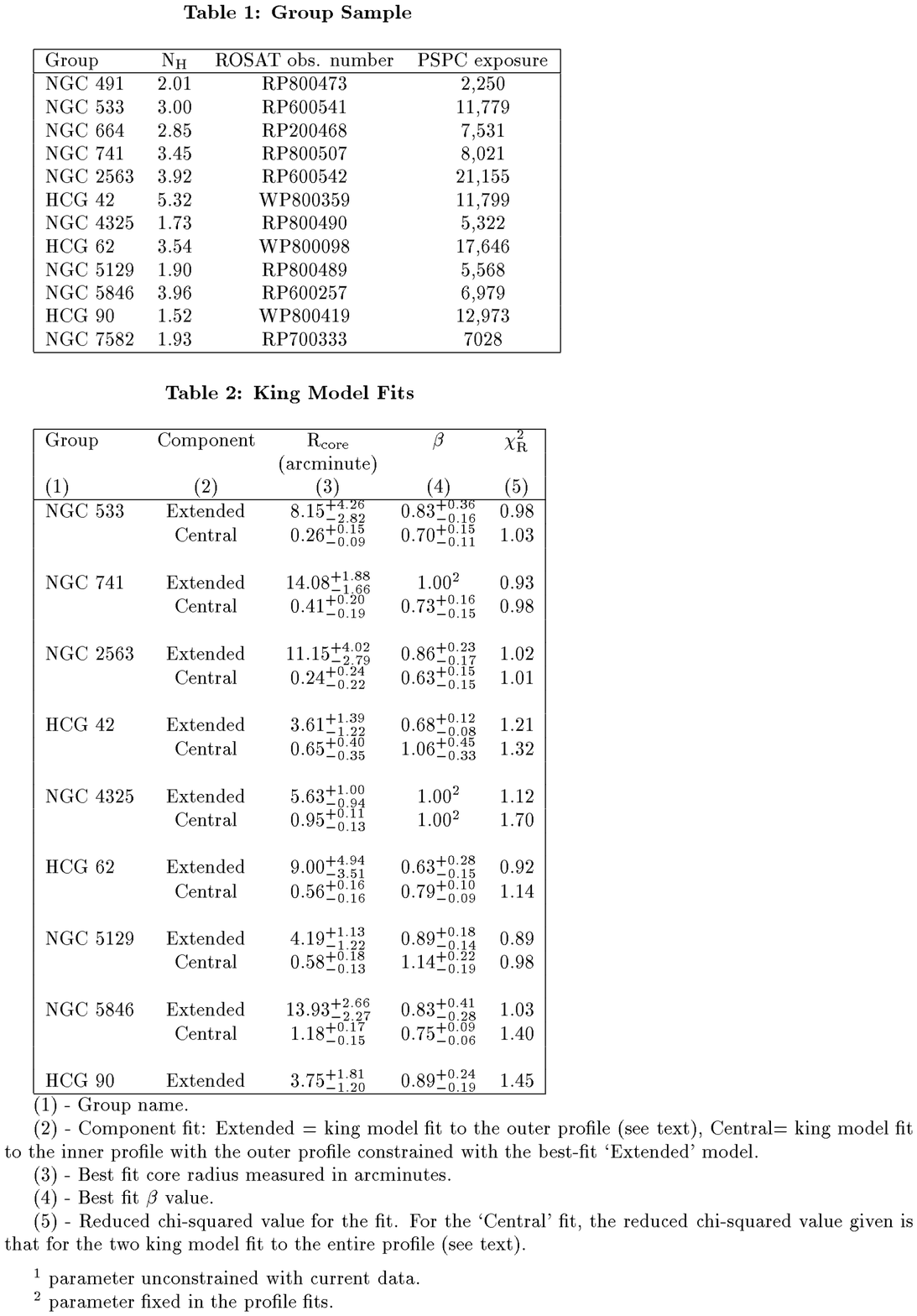}
\end{figure}
\clearpage

\begin{figure}
\plotone{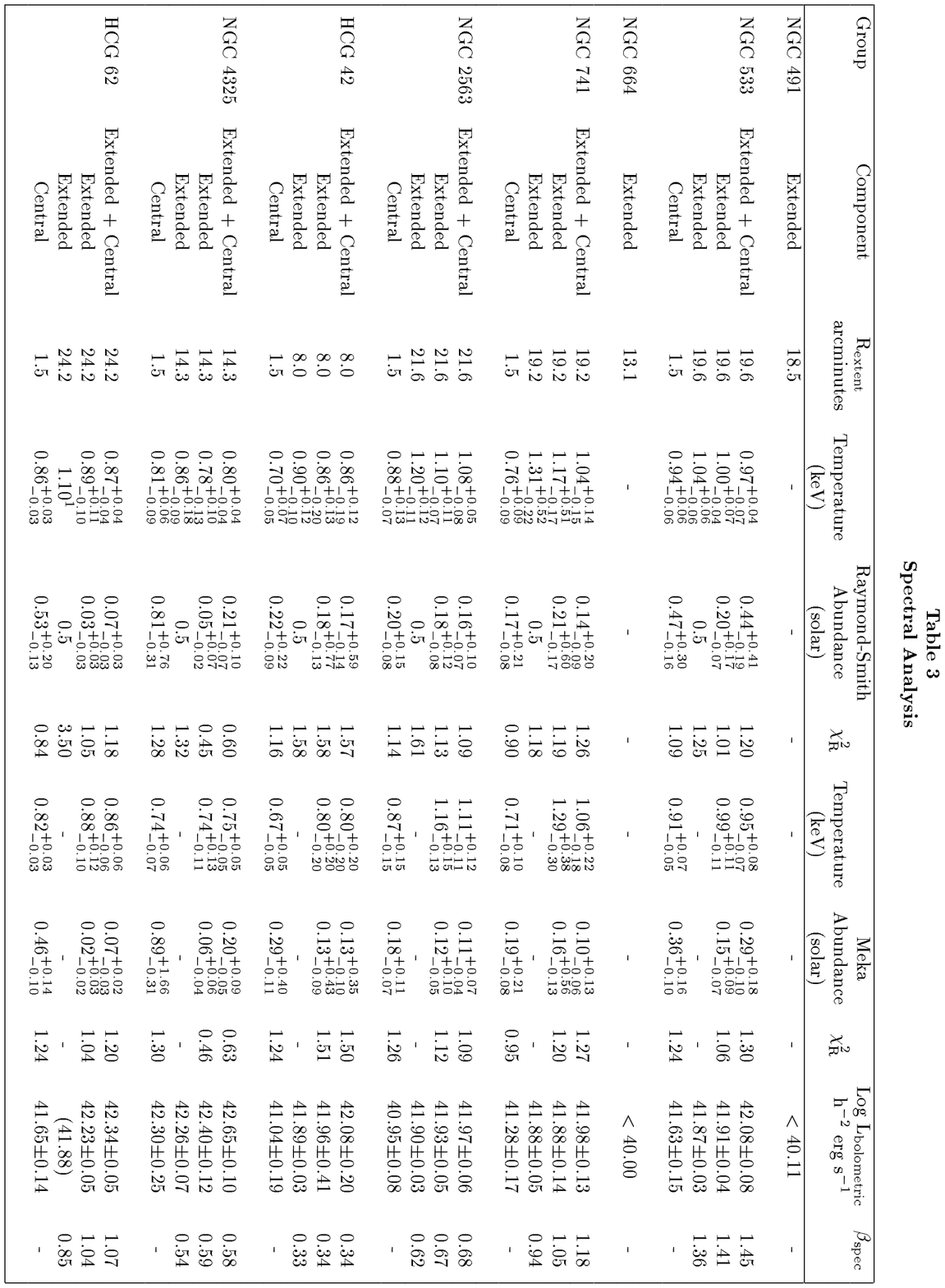}
\end{figure}
\clearpage
 
\begin{figure}
\plotone{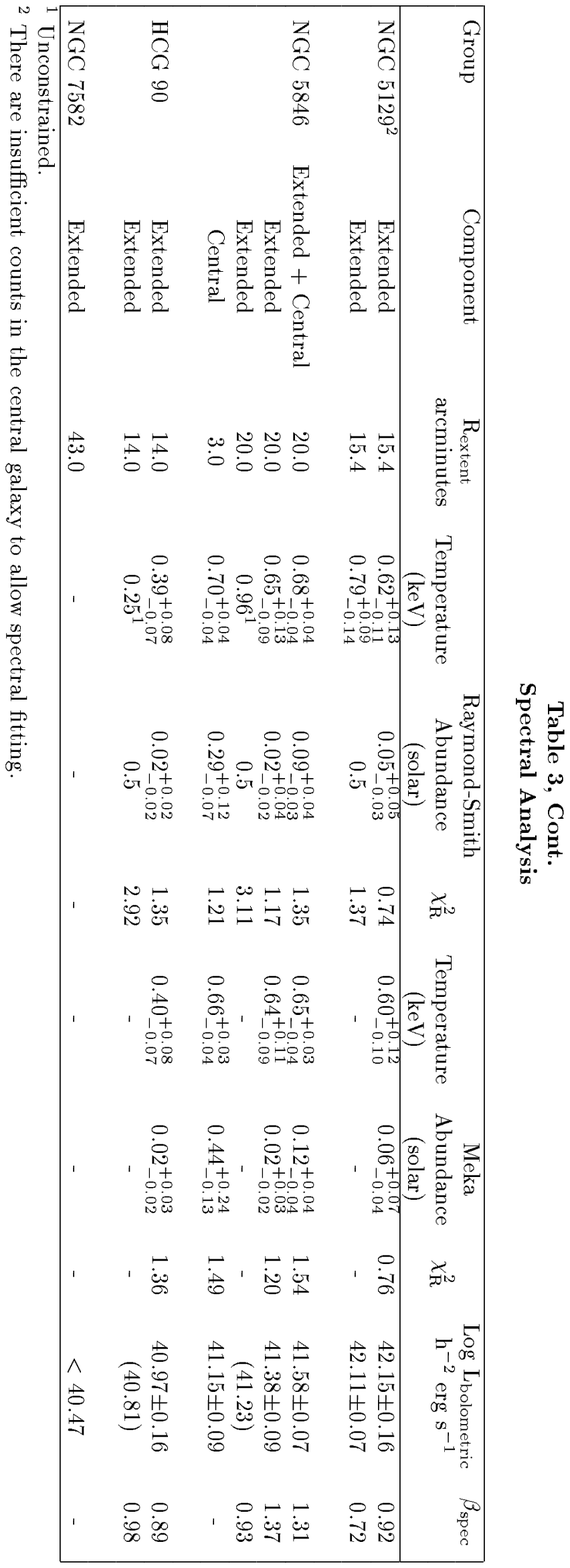}
\end{figure}
\clearpage

\begin{figure}
\plotone{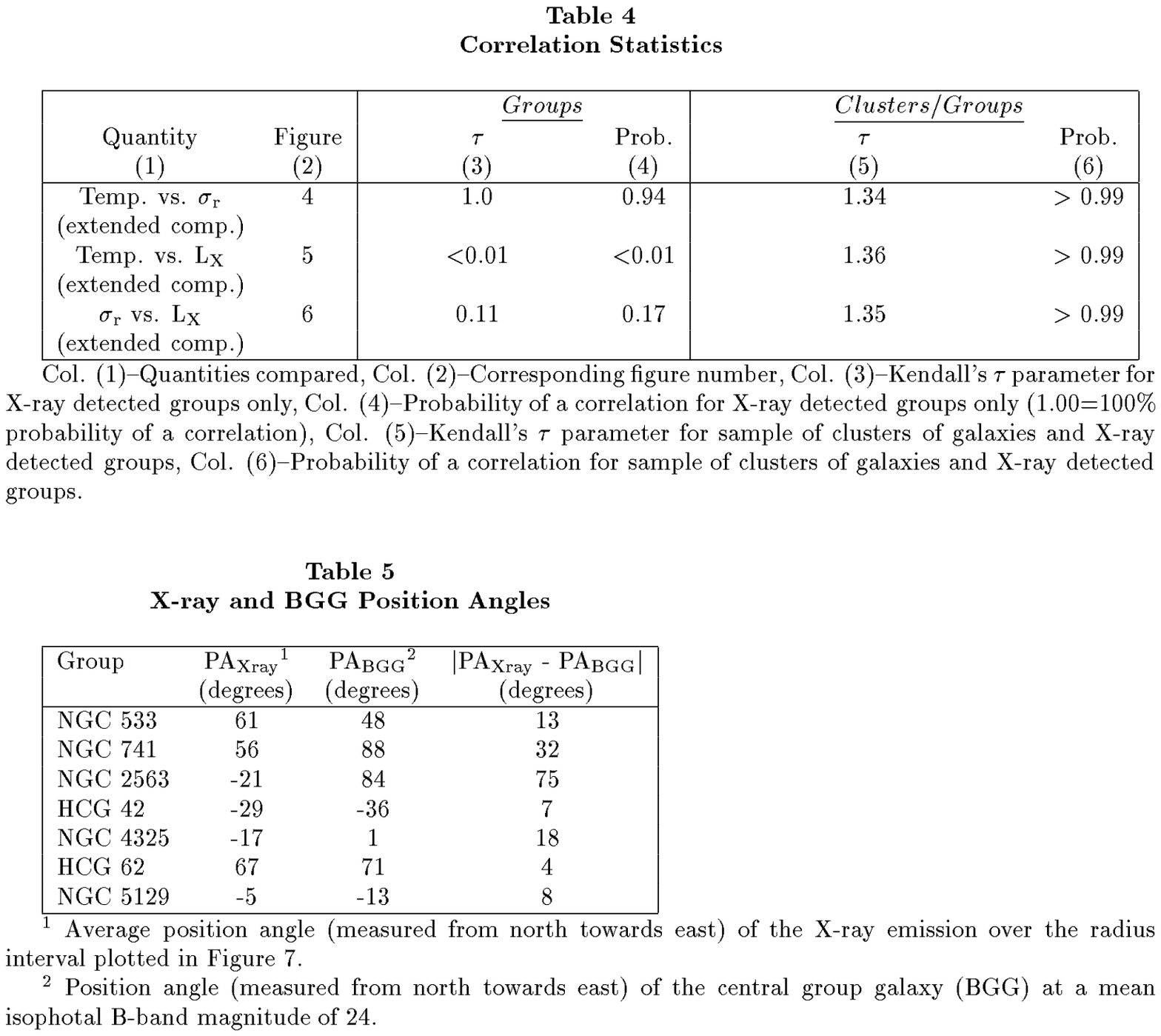}
\end{figure}
\clearpage

\clearpage

\centerline{Figure Captions}

\noindent{Fig. 1a-- Contour map of the X-ray emission in the NGC 491 group
overlayed on the STScI digitized sky survey (field of view 1$^{\rm o}$ x 1$^{\rm o}$).
No diffuse emission is detected.
The contours correspond to 2$\sigma$ and 3$\sigma$
above the background.
The data have been smoothed with a Gaussian of width 30$''$.}

\noindent{Fig. 1b-- Contour map of the diffuse emission in the NGC 533 group
overlayed on the STScI digitized sky survey (field of view 1$^{\rm o}$ x 1$^{\rm o}$). 
The emission from the 
central galaxy has not been removed.
The contours correspond to 3$\sigma$, 6$\sigma$, 9$\sigma$, 18$\sigma$,
36$\sigma$ and 72$\sigma$ above the background. 
The data have been smoothed with a Gaussian of width 30$''$.}

\noindent{Fig. 1c-- Contour map of the X-ray emission in the NGC 664 group
overlayed on the STScI digitized sky survey (field of view 1$^{\rm o}$ x 1$^{\rm o}$).
No diffuse emission is detected.
The contours correspond to 2$\sigma$ and 3$\sigma$
above the background.
The data have been smoothed with a Gaussian of width 30$''$.}

\noindent{Fig. 1d-- Contour map of the diffuse emission in the NGC 741 group
overlayed on the STScI digitized sky survey (field of view 1$^{\rm o}$ x 1$^{\rm o}$).
The emission from the
central galaxy has not been removed.
The contours correspond to 3$\sigma$, 6$\sigma$, 9$\sigma$, 18$\sigma$, 
36$\sigma$ and 72$\sigma$ above the background.
The data have been smoothed with a Gaussian of width 30$''$.}

\noindent{Fig. 1e-- Contour map of the diffuse emission in the NGC 2563 group
overlayed on the STScI digitized sky survey (field of view 1$^{\rm o}$ x 1$^{\rm o}$).
The emission from the
central galaxy has not been removed.
The contours correspond to 3$\sigma$, 6$\sigma$, 9$\sigma$, 18$\sigma$, 
36$\sigma$ and 72$\sigma$ above the background.
The data have been smoothed with a Gaussian of width 30$''$.}

\noindent{Fig. 1f-- Contour map of the diffuse emission in HCG 42 
overlayed on the STScI digitized sky survey (field of view 1$^{\rm o}$ x 1$^{\rm o}$).
The emission from the
central galaxy has not been removed.
The contours correspond to 3$\sigma$, 6$\sigma$, 9$\sigma$, 12$\sigma$, 
24$\sigma$, 60$\sigma$ and  120$\sigma$ above the background.
The data have been smoothed with a Gaussian of width 30$''$.}

\noindent{Fig. 1g-- Contour map of the diffuse emission in the NGC 4325 group
overlayed on the STScI digitized sky survey (field of view 1$^{\rm o}$ x 1$^{\rm o}$).
The emission from the
central galaxy has not been removed.
The contours correspond to 3$\sigma$, 6$\sigma$, 9$\sigma$, 18$\sigma$, 
36$\sigma$ and  72$\sigma$ above the background.
The data have been smoothed with a Gaussian of width 30$''$.}

\noindent{Fig. 1h-- Contour map of the diffuse emission in HCG 62
overlayed on the STScI digitized sky survey (field of view 1$^{\rm o}$ x 1$^{\rm o}$).
The emission from the
central galaxy has not been removed.
The contours correspond to 3$\sigma$, 6$\sigma$, 9$\sigma$, 15$\sigma$, 
30$\sigma$, 60$\sigma$ and 120$\sigma$ above the background.
The data have been smoothed with a Gaussian of width 30$''$.}

\noindent{Fig. 1i-- Contour map of the diffuse emission in the NGC 5129 group
overlayed on the STScI digitized sky survey (field of view 1$^{\rm o}$ x 1$^{\rm o}$).
The emission from the
central galaxy has not been removed.
The contours correspond to 3$\sigma$, 6$\sigma$, 9$\sigma$ and 18$\sigma$ 
above the background.
The data have been smoothed with a Gaussian of width 30$''$.}

\noindent{Fig. 1j-- Contour map of the diffuse emission in the NGC 5846 group
overlayed on the STScI digitized sky survey (field of view 1$^{\rm o}$ x 1$^{\rm o}$).
The emission from the
central galaxy has not been removed.
The contours correspond to 3$\sigma$, 6$\sigma$, 9$\sigma$, 18$\sigma$,
36$\sigma$, 72$\sigma$ and 144$\sigma$ 
above the background.
The data have been smoothed with a Gaussian of width 30$''$.}

\noindent{Fig. 1k-- Contour map of the diffuse emission in HCG 90
overlayed on the STScI digitized sky survey (field of view 1$^{\rm o}$ x 1$^{\rm o}$).
The contours correspond to 2$\sigma$, 3$\sigma$, 4$\sigma$ and 
and  5$\sigma$ above the background.
The data have been smoothed with a Gaussian of width 30$''$.}

\noindent{Fig. 1l-- Contour map of the X-ray emission in the NGC 7582 group
overlayed on the STScI digitized sky survey (field of view 1$^{\rm o}$ x 1$^{\rm o}$).
The X-ray emission $\sim$ 14$'$ north of the group center is 
coincident with the optical position of the background cluster Abell S1111 and 
presumably not related to the NGC 7582 group. The emission in the north-east
corner is associated with the rich cluster Abell 3998.
No diffuse emission from the NGC 758 group itself is detected.
The contours correspond to 3$\sigma$, 6$\sigma$, 12$\sigma$, 30$\sigma$ and
60$\sigma$
above the background.
The data have been smoothed with a Gaussian of width 30$''$.}

\noindent{Fig. 2--Surface brightness profiles for the nine X-ray detected
groups. The central galaxy emission has not been removed, except in the case of 
HCG 90. All other point sources in the field have been excluded. The 
dashed line shows the best-fit King model to the extended gas component,
while the dotted line gives the best King model to the central emission. The solid line
shows the overall two King model fit to the data. The ROSAT PSPC point spread 
function for a 1 keV source is plotted as the dotted-dashed line in the 
panel for NGC 533.}

\noindent{Fig. 3--Temperature profiles for NGC 533, NGC 2563 and HCG 62. The inner 
point in each profile corresponds to emission predominantly from the central component. Note the 
lower galaxy temperatures  in the centers of NGC 2563 and HCG 62.}

\noindent{Fig. 4--Logarithm of the X-ray temperature vs. logarithm of optical velocity 
dispersion for groups (circles) and clusters (triangles).
The cluster data are taken from Mushotzky \& Scharf (1997).
The X-ray temperature of the groups is for the extended component only.
 The straight
line gives the best-fit to the entire dataset, which corresponds to 
$\beta$ $=$ 0.99 $\pm{0.08}$.}

\noindent{Fig. 5--Logarithm of X-ray temperature vs. logarithm of
X-ray luminosity. The symbols have the same meaning as in Figure 4. 
For the groups, the luminosity plotted is for the extended component only. The total X-ray 
luminosities would typically be higher by $\sim$ 20\% if the emission 
from the central galaxy was included. The solid line gives
the best-fit to the data.}

\noindent{Fig. 6--Logarithm of velocity dispersion vs. logarithm of
X-ray temperature for the X-ray detected groups (filled circles) and clusters
(triangles). The non-X-ray detected groups are plotted using the upper
limits on L$_{\rm X}$ (arrows). As in Figure 5, the group luminosity plotted is for
the extended component only.
The solid line gives the best-fit to the data (the groups with 
upper limits are not included in the fit). The dashed line is the best-fit that
Mahdavi et al. (1997) find for their group sample (plotted over the 
velocity dispersion range in their Figure 4).}

\noindent{Fig. 7--Plots of the position angle (PA) of the X-ray emission versus radius as 
determined from ellipse fits to the isophotes of the ROSAT images. The PA is only determined
over the radius range where the 
ellipticity is greater than 0.1 (points where the ellipticity is less than 0.1 are plotted
as open squares).
The dotted line gives the average PA
of the X-ray emission over the plotted radius interval. The dashed line represents the 
optical PA of the BGG at a mean isophotal B-magnitude of 24 (see Table 5). Note the 
close alignment of the X-ray and optical PA's in many of the groups.}
\clearpage
 
\begin{figure}
\plotone{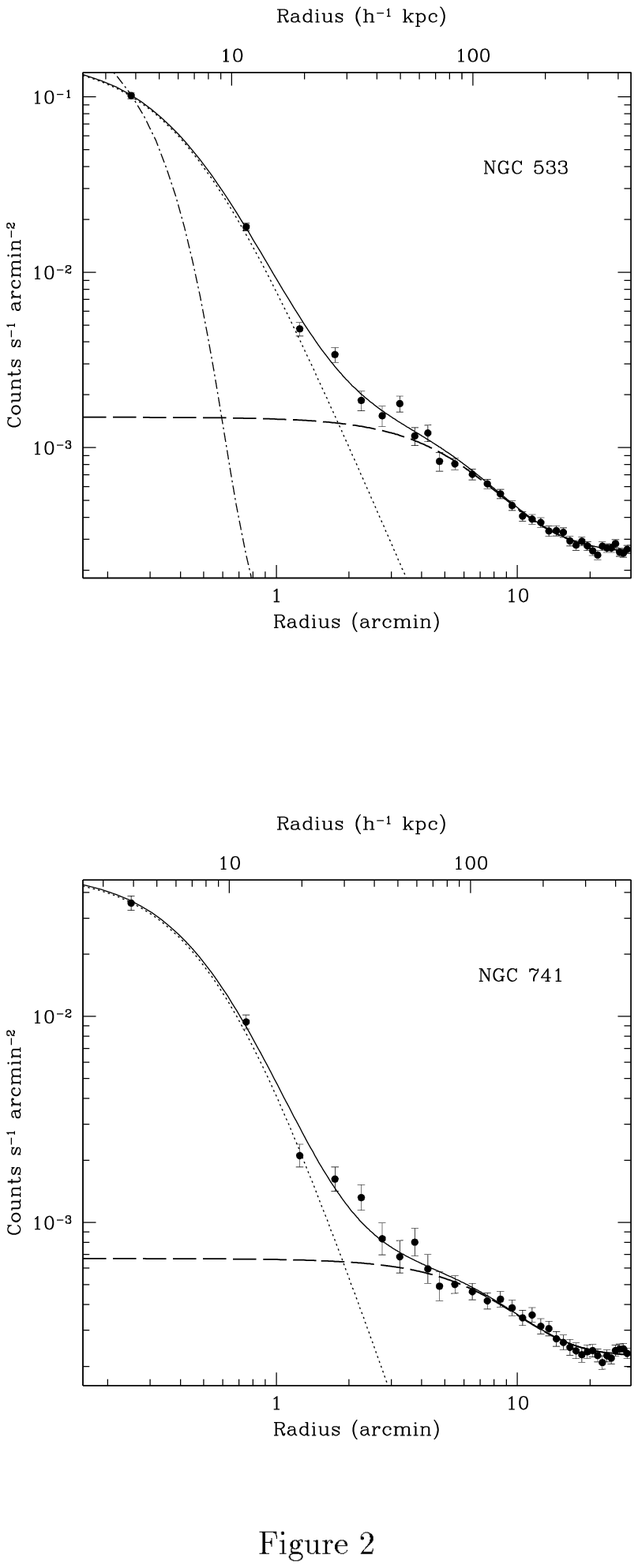}
\end{figure}
 
\clearpage
 
\begin{figure}
\plotone{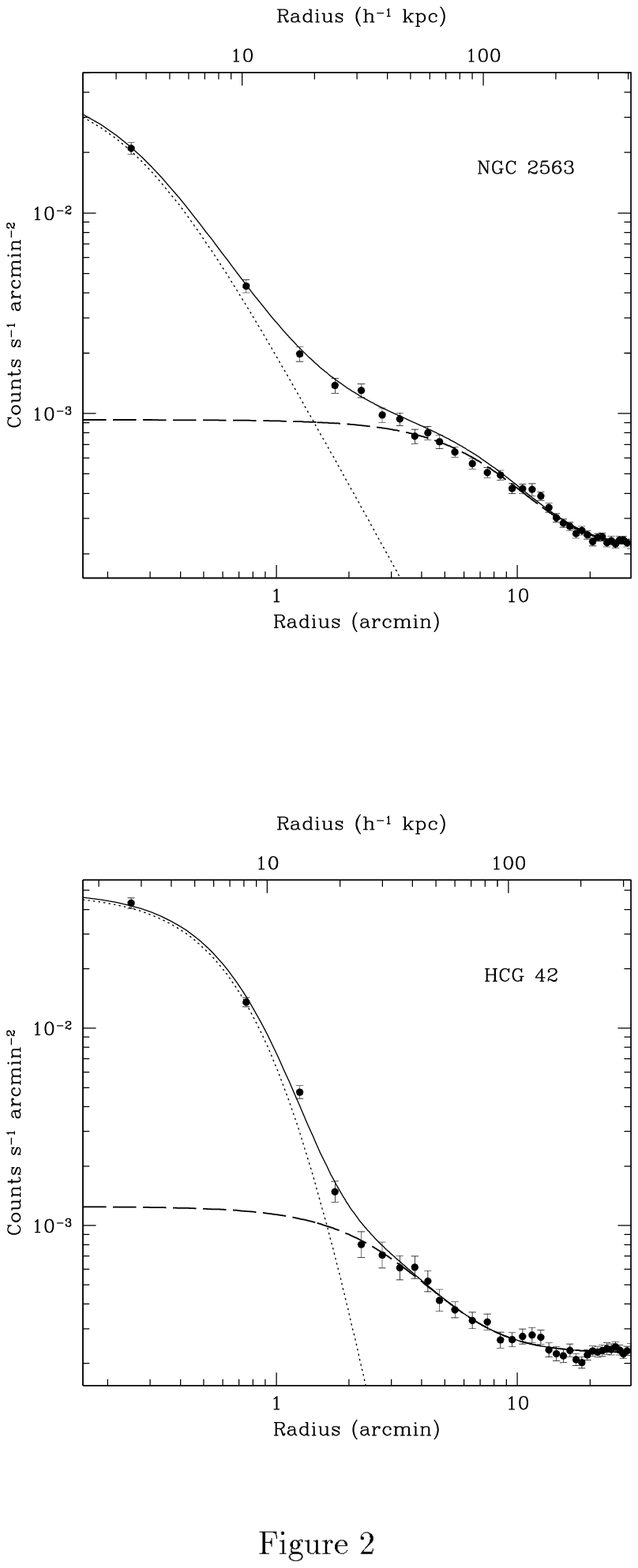}
\end{figure}
\clearpage
 
\begin{figure}
\plotone{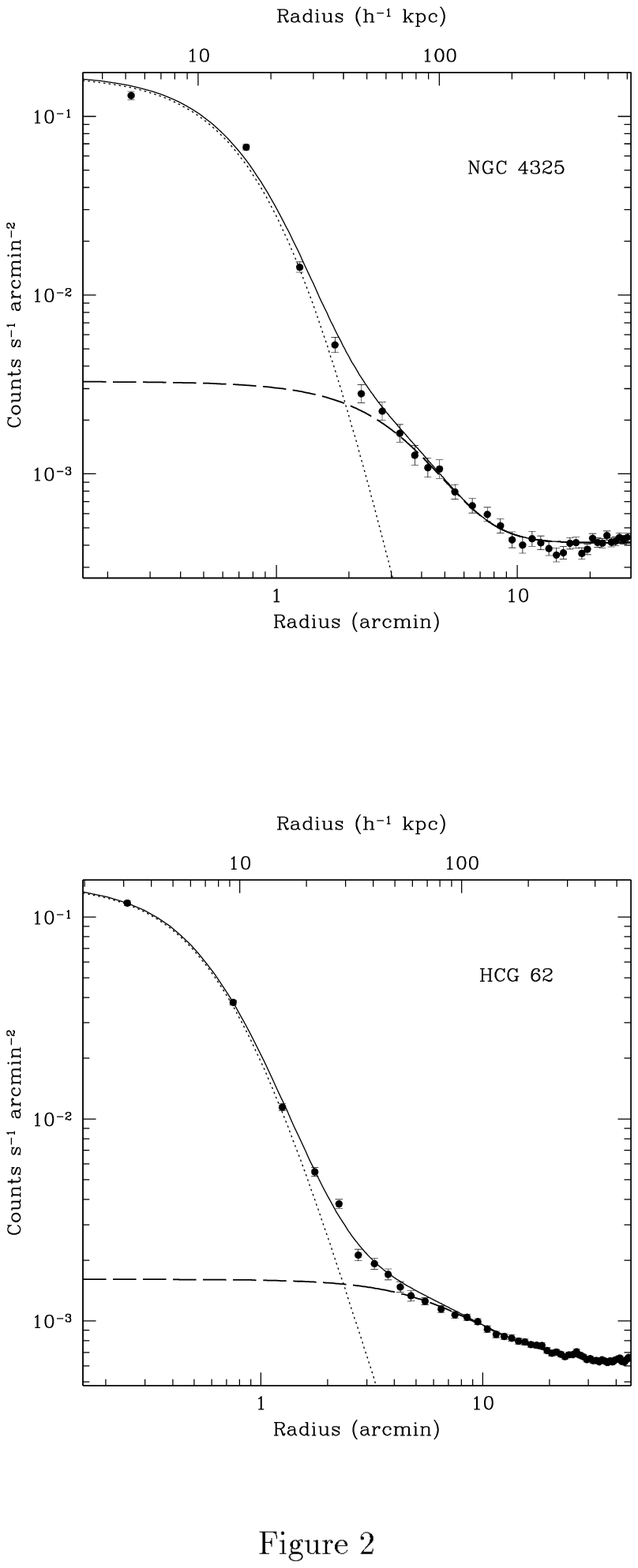}
\end{figure}
\clearpage 

\begin{figure}
\plotone{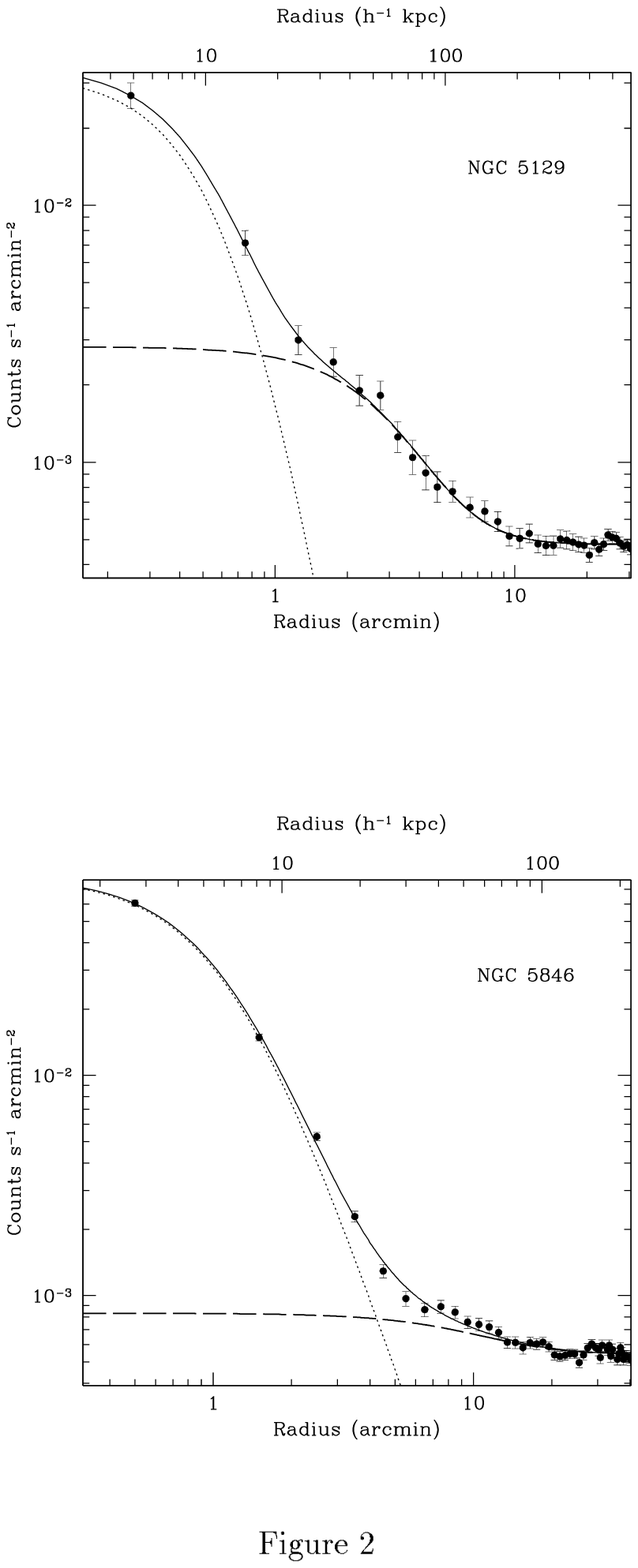}
\end{figure}
\clearpage 

\begin{figure}
\plotone{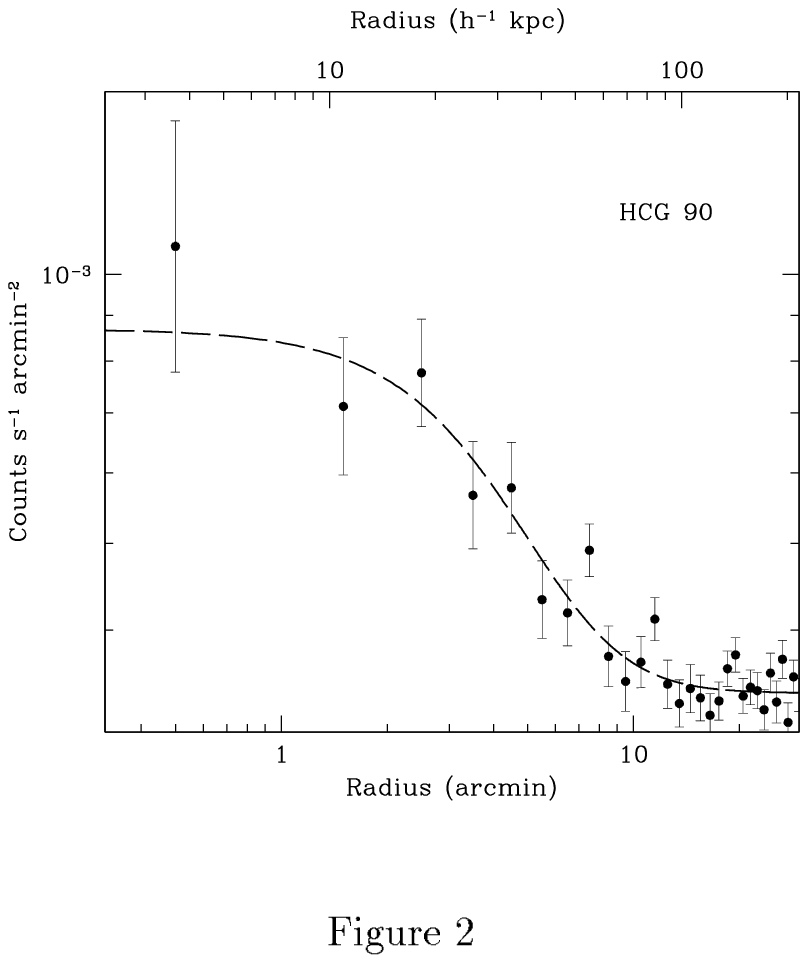}
\end{figure}
\clearpage

\begin{figure}
\plotone{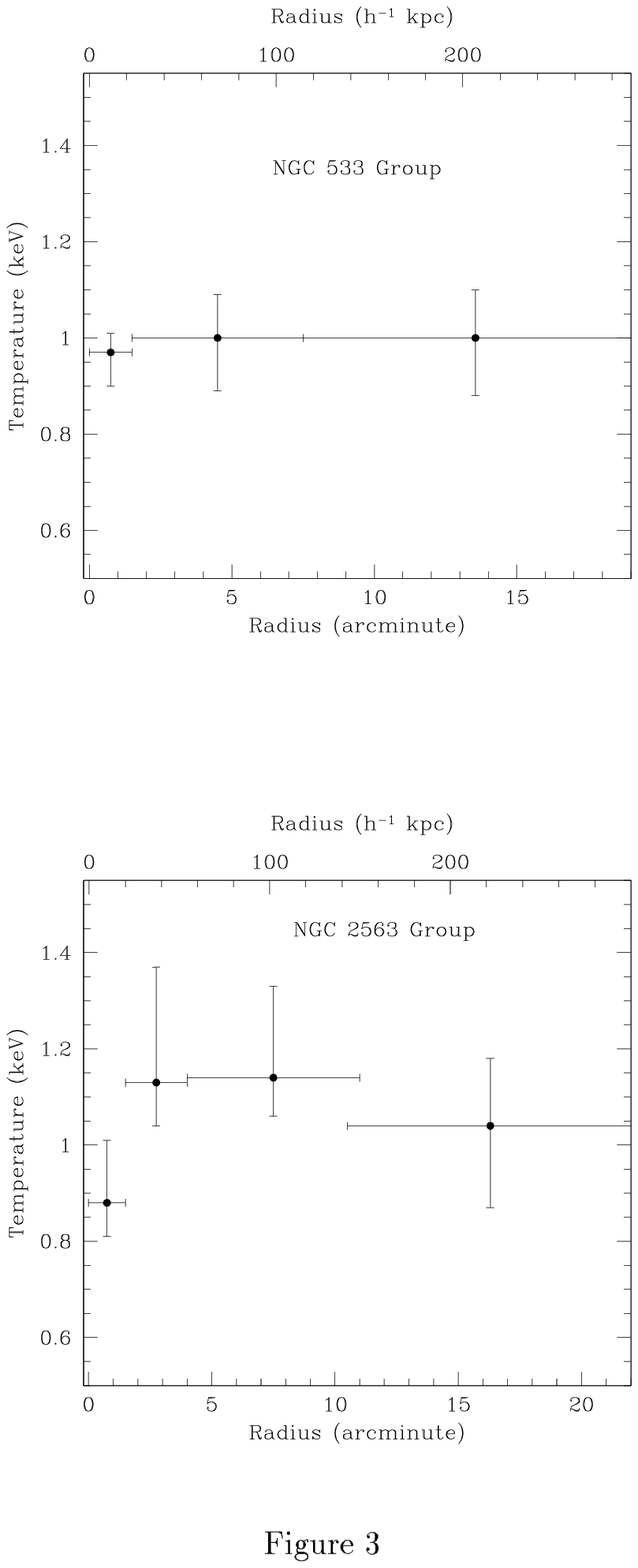}
\end{figure}
\clearpage 
 
\begin{figure}
\plotone{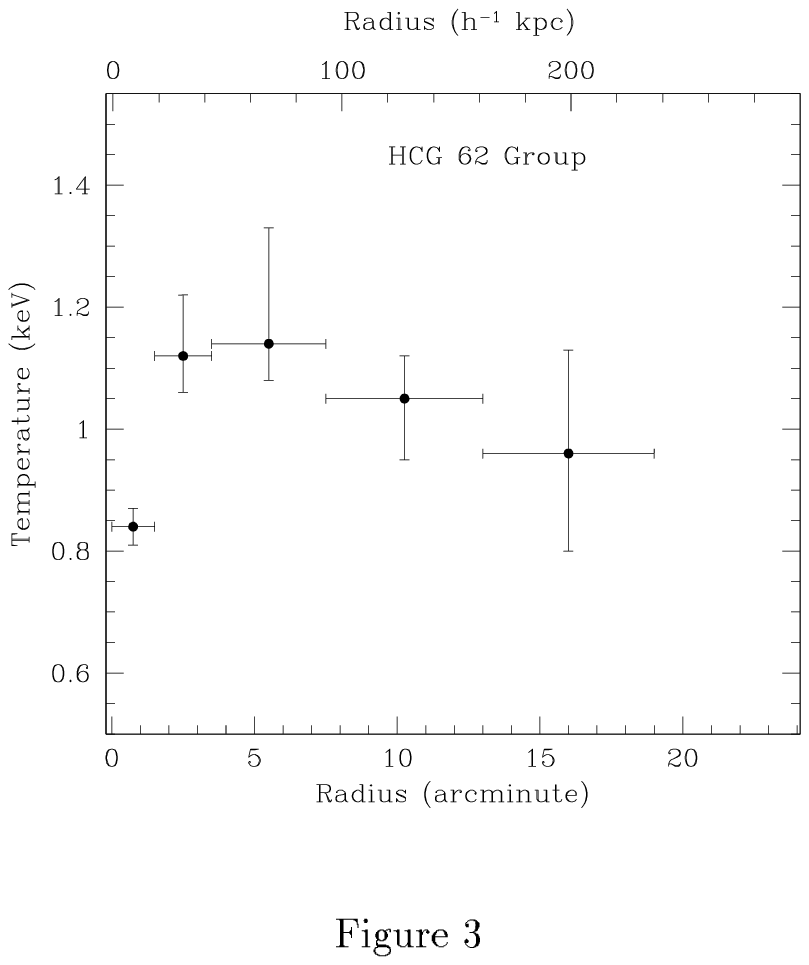}
\end{figure}
 
\clearpage

\begin{figure}
\plotone{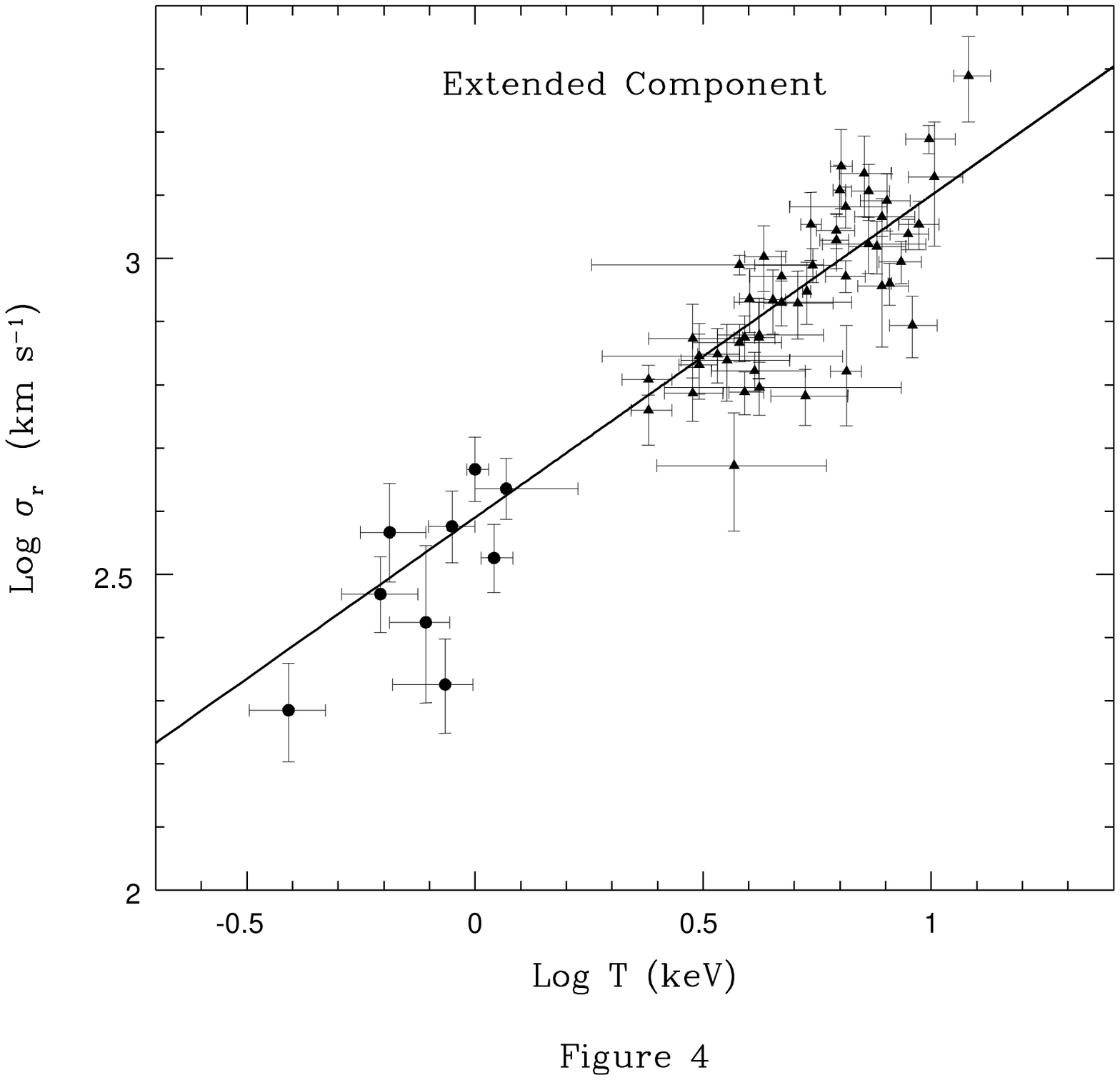}
\centerline{Figure 4}
\end{figure}

\clearpage
\begin{figure}
\plotone{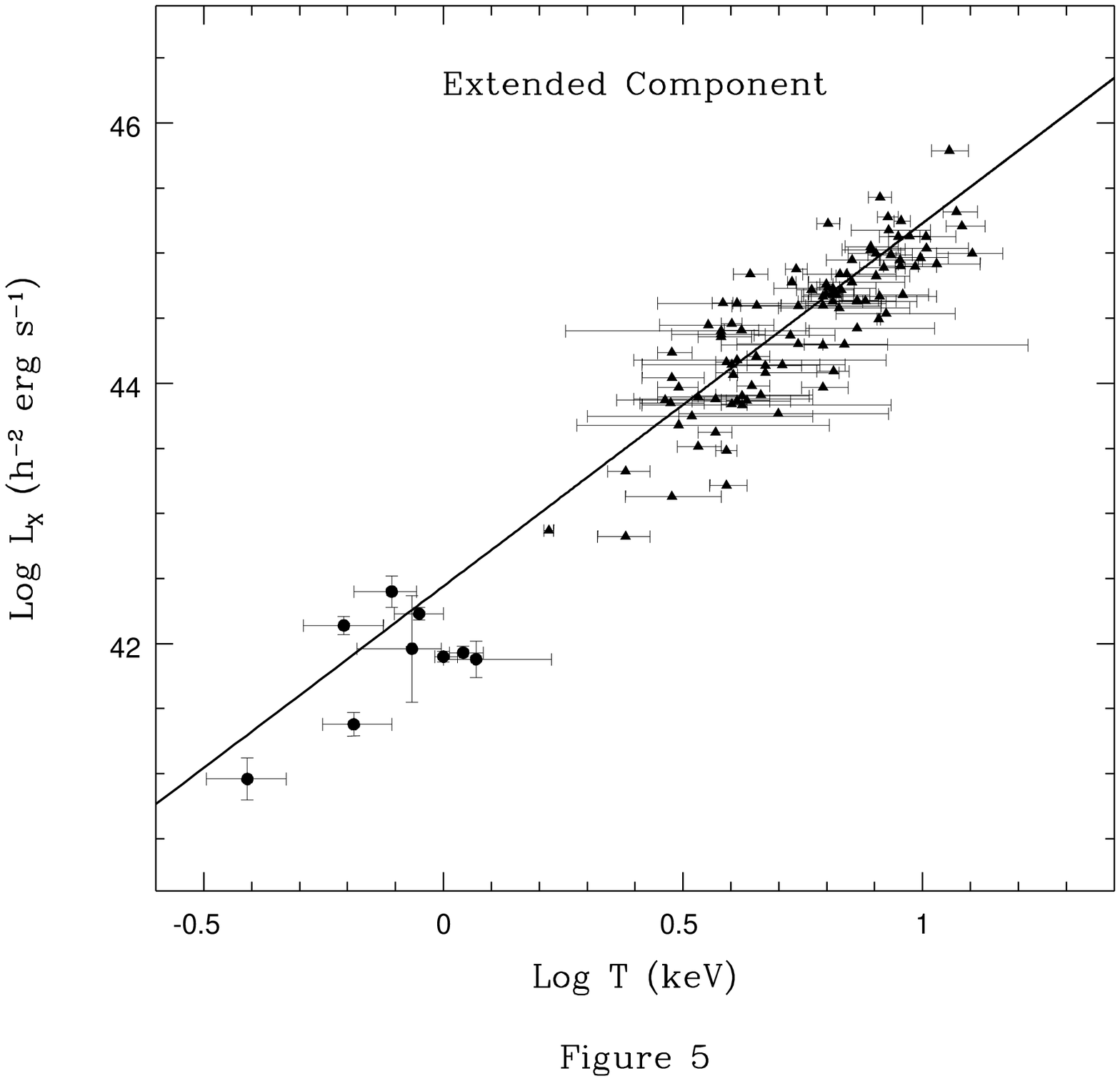}
\centerline{Figure 5}
\end{figure}

\clearpage
\begin{figure}
\plotone{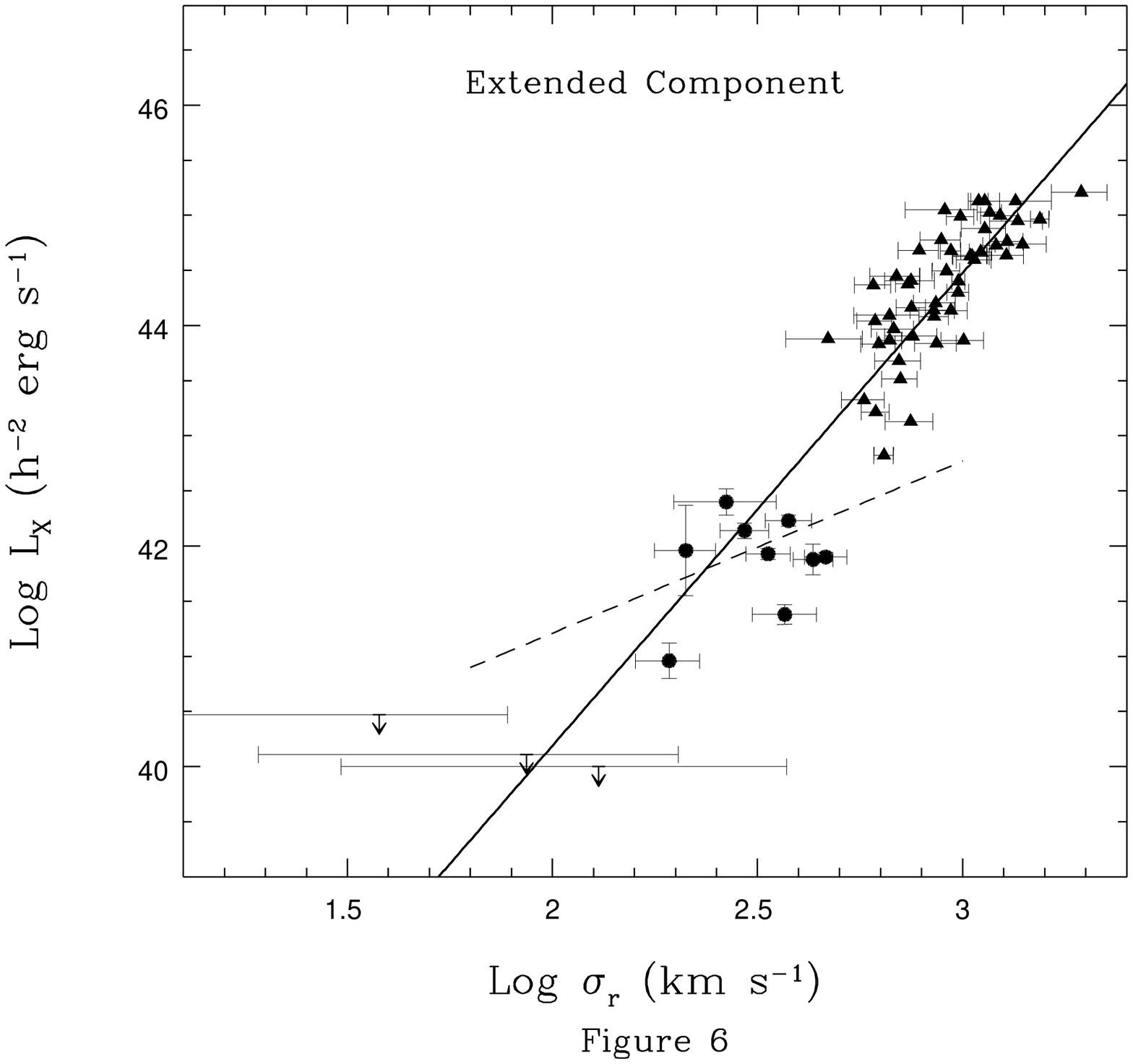}
\centerline{Figure 6}
\end{figure}

\clearpage 
\begin{figure}
\plotone{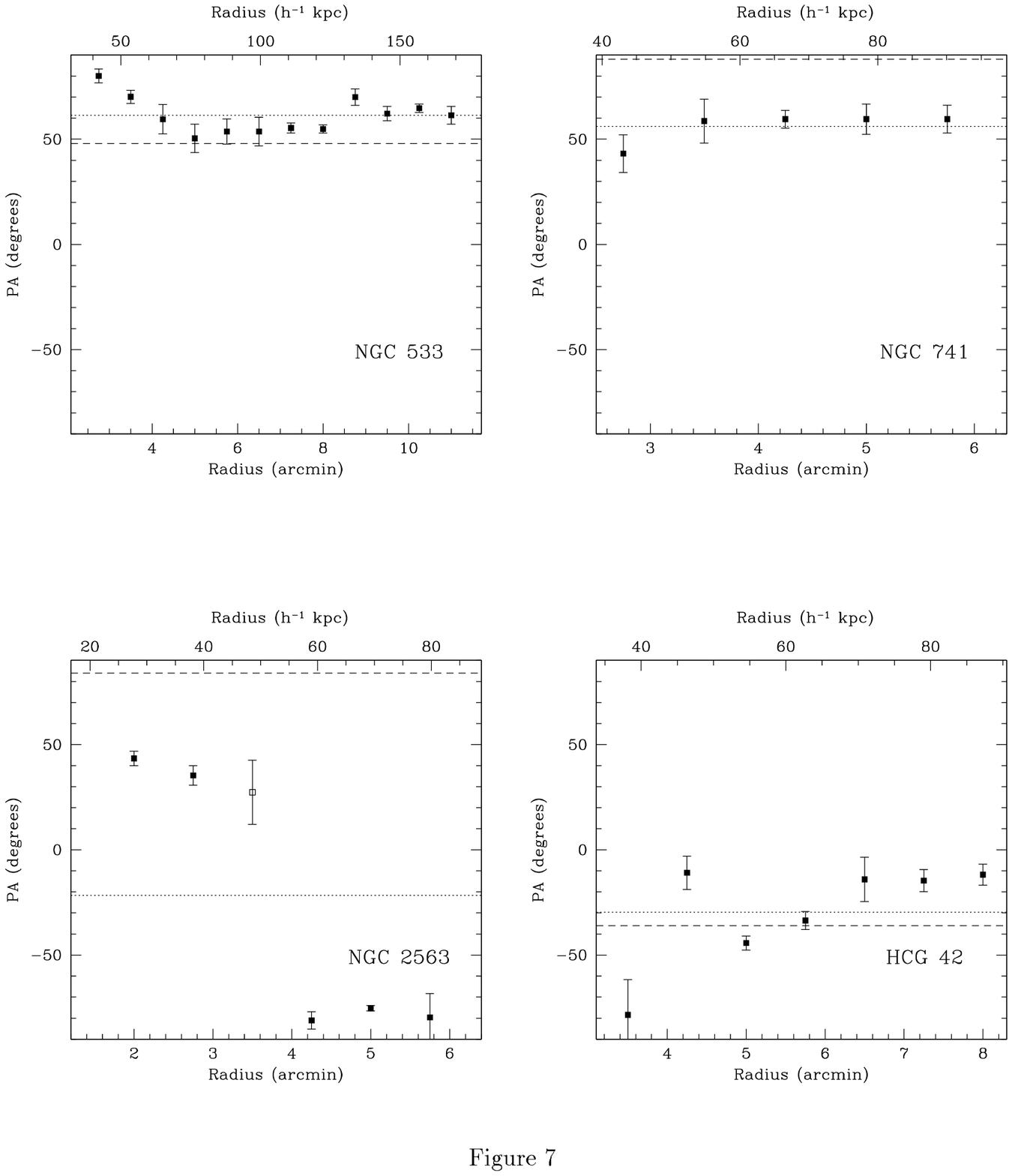}
\end{figure}
\clearpage
 
\begin{figure}
\plotone{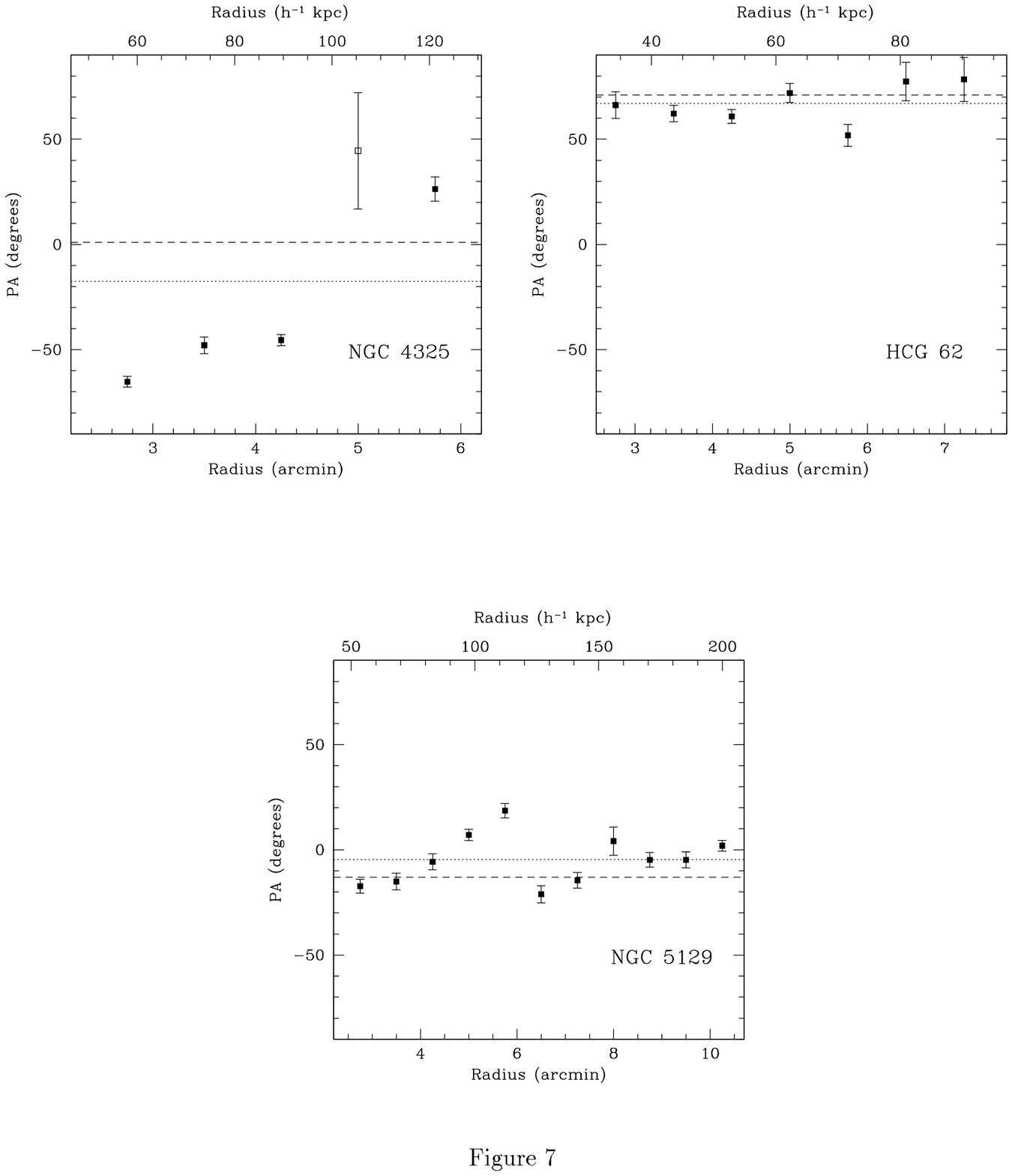}
\end{figure}
\clearpage


\begin{references}

\hi{Allen, S. W., Fabian, A. C., Edge, A. C., Bohringer, H.,
White, D. A. 1995, MNRAS, 275, 741}

\hi{Bauer, F., \& Bregman, J. N. 1996, ApJ, 457, 382}

\hi{Buote, D. A., \& Canizares, C. R. 1994, \apj, 427, 86}

\hi{Burns, J. O., Ledlow, M. J., Loken, C., Klypin, A., Voges, W.,
Bryan, G. L., Norman, M. L., \& White, R. A 1996, \apj, 467, L49.}

\hi{Canizares, C. R., Fabbiano, G., \& Trinchieri, G. 1987, \apj, 312, 503}

\hi{Crawford, C. S., \& Fabian, A. C. 1992, MNRAS, 259, 265}

\hi{David, L. P., Slyz, A., Jones, C., Forman, W., Vrtilek, S. D., \&
Arnaud, K. A., 1993, \apj, 412, 479}
 
\hi{David, L. P., Jones, C., Forman, W., \& Daines, S. 1994, ApJ, 428, 544}
 
\hi{Davis, D. S., Mulchaey, J. S., Mushotzky, R. F., \& Burstein, D. 1996,
ApJ, 460, 601}

\hi{Dell'Antonio, I. P., Geller, M. J., \& Fabricant, D. G. 1994, AJ, 107, 427}

\hi{Diaferio, A., Geller, M. J., \& Ramella, M. 1995, AJ, 109, 2293}

\hi{Doe, S. M., Ledlow, M. L., Burns, J. O., \& White, R. A. 1995, AJ, 110, 46}
 
\hi{Ebeling, H, Voges, W., \& B$\ddot{o}$hringer, H. 1994, ApJ, 436, 44}

\hi{Edge, A. C., \& Stewart, G. C. 1991, MNRAS, 252, 428}

\hi{Eskridge, P. B., Fabbiano, G., \& Kim, D.-W. 1995, \apjs, 97, 141}

\hi{Fabbiano, G., Kim, D.-W., \& Trinchieri, G. 1992, \apjs, 80, 531}

\hi{Fabian, A. C. 1994, ARAA, 32, 277}

\hi{Forman, W., Jones, C., \& Tucker, W. 1985, \apj, 293, 102}

\hi{Hasinger, G., Turner, T. J., George, I. M., \& Boese, G. 1992,
NASA/GSFL Office of Guest Investigator Programs, Calibration Memo 
CAL/ROS/92-001} 

\hi{Heckman, T. M., Baum, S. A., van Breugel, W. J. M, \& McCarthy, P. 1989,
\apj, 338, 48}

\hi{Henry, J. P. et al. 1995, ApJ, 449, 422}

\hi{Hernquist, L., Katz, N., \& Weinberg, D. H. 1995, \apj, 442, 57}

\hi{Ikebe, Y. et al. 1996, Nature, 379, 427}

\hi{Lubin, L. M., \& Bahcall, N. A. 1993, \apjl, 415, L17}

\hi{Mahdavi, A., Bohringer, H., Geller, M. J., \& Ramella, M. 1997, \apj, in press}

\hi{Makishima, K. 1995, in {\it Dark Matter}, American Institute of Physics,
New York, ed. S. Holt \& C. L. Bennett, p. 172}

\hi{Matsumoto, H. et al. 1997, \apj, in press}

\hi{Mewe, R., Gronenschild, E. H., \& van den Oord, G. H. J. 1985, AA Supl., 62, 197} 

\hi{Merritt, D. 1985, \apj, 289, 18}

\hi{Mulchaey, J. S., Davis, D. S., Mushotzky, R. F., \& Burstein, D. 1993,
ApJ, 404, L9}
 
\hi{Mulchaey, J. S., Davis, D. S., Mushotzky, R. F., \& Burstein, D. 1996a,
ApJ, 456, 80}

\hi{Mulchaey, J. S., Mushotzky, R. F., Burstein, D., \& Davis, D. S. 1996b,
ApJ, 456, L5}

\hi{Mushotzky, R. F., \& Scharf, C. A. 1997, \apjl, in press}

\hi{Pildis, R. A., Bregman, J. N., \& Evrard, A. E. 1995, ApJ, 443, 514}
 
\hi{Ponman, T. J., \& Bertram, D. 1993, Nature, 363, 51}

\hi{Ponman, T. J., Bourner, P. D. J., Ebeling, H., \& Bohringer, H. 1996, 
MNRAS, 283, 690}

\hi{Ramella, M., Geller, M. J., \& Huchra, J. P. 1989, \apj, 344, 57}

\hi{Rhee, G., Van Haarlem, M., \& Katgert, P. 1992, AJ, 103, 1721}

\hi{Sarazin, C. L., Burns, J. O., Roettiger, K., \& Mcnamara, B. R. 1995,
\apj, 447, 559}

\hi{Snowden, S. L., Mccammon, D., Burrows, D. N., \&
Mendehall, J. A. 1994, \apj, 424, 714}

\hi{Stark, A. A., Gammie, C. F., Wilson, R. W., Bally, J., 
Linke, R. A., Heiles, C., \& Hurwitz, M. 1992, \apjs, 79, 77}

\hi{Thomas, P. A., Fabian, A. C., Arnaud, K. A., 
Forman, W., \& Jones, C. 1986, MNRAS, 222, 655}

\hi{Trinchieri, G., Fabbiano, G., \& Canizares, C. R. 1986, \apj, 310, 637}

\hi{Trinchieri, G., Fabbiano, G., \& Kim, D.-W. 1997, AA , 318, 361}

\hi{Zabludoff, A. I, \& Mulchaey, J. S. 1997, \apj, submitted (Paper I)}

\end{references}
\end{document}